\documentclass[aps,twocolumn,prx,preprintnumbers,showpacs,amsmath,amssymb,superscriptaddress]{revtex4-1}
\usepackage[dvipdfm, colorlinks, citecolor=blue]{hyperref} % this is for eps format
\usepackage{graphicx}
\usepackage{dcolumn}
\usepackage{threeparttable}
\usepackage{multirow}
\usepackage{booktabs}
\usepackage{txfonts}
\usepackage{xcolor}
\usepackage{bm}
\usepackage{amssymb}
\usepackage{amsmath}
\usepackage{latexsym}
\usepackage{epsfig}
\usepackage{amsbsy}
\usepackage{array}
\usepackage{tabularx}
\usepackage{esvect} %vector
\usepackage{extarrows}

\begin{document}

\title{Comparative \emph{ab initio} study of the structural, electronic, magnetic, and dynamical properties of \\
LiOsO$_3$ and  NaOsO$_3$}

\author{Peitao Liu}
\email{peitao.liu@univie.ac.at}
\affiliation{University of Vienna, Faculty of Physics and Center for
Computational Materials Science, Sensengasse 8, A-1090, Vienna,
Austria}

\author{Jiangang He}
\affiliation{Department of Materials Science and Engineering,
Northwestern University, Evanston, Illinois 60208, United States}

\author{Bongjae Kim}
\affiliation{Department of Physics, Kunsan National University, Gunsan 54150, Korea}

\author{Sergii Khmelevskyi}
\affiliation{University of Vienna, Faculty of Physics and Center for
Computational Materials Science, Sensengasse 8, A-1090, Vienna,
Austria}

\author{Alessandro Toschi}
\affiliation{Institute for Solid State Physics, Vienna University of Technology, 1040 Vienna, Austria}

\author{Georg Kresse}
\affiliation{University of Vienna, Faculty of Physics and Center for
Computational Materials Science, Sensengasse 8, A-1090, Vienna,
Austria}

\author{Cesare Franchini}
\affiliation{University of Vienna, Faculty of Physics and Center for
Computational Materials Science, Sensengasse 8, A-1090, Vienna,
Austria}
\affiliation{Dipartimento di Fisica e Astronomia, Universit\`{a} di Bologna, 40127
Bologna, Italy}

\begin{abstract}
Despite similar chemical compositions, LiOsO$_3$ and NaOsO$_3$ exhibit remarkably distinct structural, electronic, magnetic, and spectroscopic properties.
At low temperature, LiOsO$_3$ is a polar bad metal with a rhombohedral $R3c$ structure without the presence of long-range magnetic order,
whereas NaOsO$_3$ is a $G$-type antiferromagnetic insulator with an orthorhombic $Pnma$ structure.
By means of comparative first-principles DFT+$U$ calculations with the inclusion of the spin-orbit coupling,
we ($i$) identify the origin of the different structural ($R3c$ vs. $Pnma$) properties using a symmetry-adapted soft mode analysis,
($ii$) provide evidence that all considered exchange-correlation functionals (LDA, PBE, PBEsol, SCAN, and HSE06)
and the spin disordered polymorphous descriptions
are unsatisfactory to  accurately describe the electronic and magnetic properties of both systems simultaneously,
and ($iii$) clarify that the distinct electronic (metallic vs. insulating) properties originates mainly from a cooperative steric and magnetic effect.
Finally, we find that although at ambient pressure LiOsO$_3$ with a $Pnma$ symmetry and
NaOsO$_3$ with a $R\bar{3}c$ symmetry are energetically unfavorable, they do not show soft phonons and therefore are
dynamically stable. A pressure-induced structural phase transition from $R3c$ to $Pnma$ for LiOsO$_3$ is predicted,
whereas for NaOsO$_3$ no symmetry change is discerned in the considered pressure range.
\end{abstract}

\maketitle

%--------------------------------------------------------------------------------
\section{Introduction}\label{sec:intro}
%--------------------------------------------------------------------------------

Transition-metal oxide (TMO) perovskites represent a rich ground for the emergence of intriguing properties and novel phases originating
from the complex interplay of different interactions with the cross coupling of spin, charge, orbital, and lattice degrees of
freedom~\cite{doi:10.1146/annurev-conmatphys-020911-125138,Martins_2017}.
When the transition-metal elements shift from 3$d$ to 5$d$,  spin-orbit coupling (SOC) is
enhanced owning to the increased atomic mass and correlation effects weaken due to
 the extended nature of 5$d$ orbitals and associated widening of the band width~\cite{PhysRevMaterials.2.024601}.
Their comparable strength and cooperative interplay in 5$d$ TMOs
give rise to, e.g.,  a novel $J_{\rm eff}=1/2$ Mott-insulating state in an otherwise
metallic Sr$_2$IrO$_4$~\cite{PhysRevLett.101.076402,PhysRevLett.102.017205,PhysRevB.92.054428}.
In addition to iridates, osmium TMOs have also stimulated a lot of interest~\cite{YAMAURA201645}, e.g.,  because of
the observed unusual ferroelectric-like structural transition in metallic LiOsO$_3$~\cite{Shi2013, PhysRevLett.122.227601} and
continuous metal-insulator transition (MIT)~\cite{PhysRevB.80.161104, PhysRevB.94.241113}
and anomalously strong spin-phonon-electronic coupling~\cite{Calder2015} in NaOsO$_3$.

Despite similar chemical compositions, same electronic configurations ($5d^3$),
and comparable electronic correlation and SOC strengths,
LiOsO$_3$ and NaOsO$_3$ exhibit strikingly different structural, electronic, and magnetic properties.
Experimentally, LiOsO$_3$ displays a bad metallic character over the whole temperature range~\cite{Shi2013}.
It possesses a centrosymmetric $R\bar{3}c$ rhombohedral structure at high temperature
and undergoes a second-order ferroelectric-like structural transition to a noncentrosymmetric
$R3c$ structure at $T_s$ = 140 K~\cite{Shi2013}.  The origin of this transition
was understood by the instability of Li ions along the polar axis and the incomplete
screening of the short-range dipole-dipole interactions~\cite{PhysRevB.90.094108, PhysRevB.89.201107, PhysRevB.90.195113,PhysRevB.91.064104}.
Although a Curie-Weiss-like behaviour is observed below $T_s$,
no evidence of long-range magnetic order is found even down to very low temperature~\cite{Shi2013,doi:10.7566/JPSCP.21.011013}.
By contrast, NaOsO$_3$ displays an orthorhombic $Pnma$ structure and undergoes a continuous second-order MIT~\cite{PhysRevB.80.161104}
with a small optical gap ($\sim$ 0.1 eV)~\cite{LoVecchio2013}, which is accompanied by the onset of a long-range $G$-type antiferromagnetic (AFM)
ordering at a N\'{e}el temperature $T_N$ = 410~K with a magnetic moment of 1.0 $\mu_B$~\cite{PhysRevLett.108.257209}.
The MIT in NaOsO$_3$ was initially explained by a Slater mechanism~\cite{PhysRevB.80.161104, PhysRevLett.108.257209, LoVecchio2013,PhysRevB.85.174424,PhysRevB.87.115119}
and later better interpreted in terms of a continuous Lifshitz-type transition driven by magnetic fluctuations~\cite{PhysRevB.94.241113, PhysRevLett.120.227203,PhysRevB.97.184429}.
A detailed comparison between characteristic ground-state (GS) properties and energy scales of LiOsO$_3$ and NaOsO$_3$ is summarized in Table~\ref{Table: LOO_vs_NOO}.

\begin{table}[h!]
\caption {Collection of the low-temperature GS properties of LiOsO$_3$ and NaOsO$_3$.
The $t_{\rm 2g}$ bandwidth and orbital-averaged Coulomb repulsion $U$ and the Hund's coupling $J$ are calculated by LDA and the constrained random phase approximation (cRPA)~\cite{PhysRevB.74.125106}. For LiOsO$_3$ there is no indication of a magnetic ordering~\cite{Shi2013}, though
the susceptibility shows the Curie-Weiss-like behaviour suggesting the presence of localised paramagnetic (PM) moments~\cite{doi:10.7566/JPSCP.21.011013}.
}
\begin{ruledtabular}
\begin{tabular}{lcc}
    & LiOsO$_3$ & NaOsO$_3$ \\
\hline
Electronic configuration & Os$^{5+}$ ($t^3_{\rm 2g}$)   & Os$^{5+}$ ($t^3_{\rm 2g}$)  \\
Crystal symmetry  & $R3c$~\cite{Shi2013}   & $Pnma$~\cite{PhysRevB.80.161104} \\
Modes condensation from $Pm\bar{3}m$  & $R^-_5$, $\Gamma^-_4$  & $R^-_5$, $M^+_2$ \\
Goldschmidt tolerance factor $t$ & 0.75   & 0.84  \\
Ionic radius ($\AA$) of Li$^{+}$ (Na$^{+})$                    &   0.90      &   1.16    \\
Experimental volume ($\AA^3$/f.u.)   & 48.65~\cite{Shi2013} & 54.37~\cite{PhysRevB.80.161104}\\
Averaged Os-O bond length (\AA)  & 1.944 & 1.941 \\
Band gap (eV)   & Metal   & 0.1~\cite{LoVecchio2013}  \\
Magnetic order & PM~\cite{Shi2013,doi:10.7566/JPSCP.21.011013} & $G$-AFM~\cite{PhysRevLett.108.257209} \\
Local magnetic moment  ($\mu_B$)  & ---       & 1.0~\cite{PhysRevLett.108.257209} \\
SOC strength $\lambda$ (eV/Os)   & 0.3~\cite{Daniel2019} & 0.3~\cite{Daniel2019} \\
$t_{\rm 2g}$ bandwidth (no SOC) (eV)  & 3.47 & 3.93  \\
$t_{\rm 2g}$ bandwidth (with SOC) (eV)  & 3.63 & 4.06 \\
Orbital-averaged $U^{\rm cRPA}_{\rm noSOC}$ (eV)   & 1.94  &  1.86  \\
Orbital-averaged $J^{\rm cRPA}_{\rm noSOC}$ (eV)    & 0.25  & 0.24 \\
\end{tabular}
\end{ruledtabular}
\label{Table: LOO_vs_NOO}
\end{table}

In addition, the two compounds exhibit remarkably distinct spectroscopic properties~\cite{LoVecchio2013,PhysRevB.93.161113}.
Upon raising the temperature, the AFM insulating state in NaOsO$_3$ develops into a bad metal (pseudogap) regime, which
is transformed into a paramagnetic (PM) metallic phase with relatively good Fermi liquid properties at high temperature,
as revealed by terahertz and infrared spectroscopy~\cite{LoVecchio2013}.
Conversely, the optical spectrum of LiOsO$_3$ rapidly loses the sign of metallic coherence as the temperature increases.
At room temperature, the Drude peak is replaced by a slight low-frequency downturn~\cite{PhysRevB.93.161113},
similar to the behavior observed in undoped V$_2$O$_3$, a prototypical material on the verge of a Mott MIT~\cite{PhysRevB.77.113107}.
By conducting a first-principle many-body analysis we have demonstrated that the distinct high-temperature spectroscopic properties of these
two compounds originate from their different degrees of proximity to an adjacent Hund's-Mott insulating phase~\cite{Daniel2019}.

In this paper, by conducting a variety of comparative computational experiments rooted in density functional theory (DFT)
plus an on-site Hubbard $U$ and SOC effects, we aim to cast some light on the the origin of the
different low-temperature GS properties of NaOsO$_3$ and LiOsO$_3$.

At first, using the symmetry-adapted soft mode analysis we clarify the structural differences
by identifying the symmetry path from the ideal cubic perovskite structure to the $R3c$ (LiOsO$_3$) and  $Pnma$ (NaOsO$_3$) phases.
Then, by applying a wide variety of DFT functionals (local, semilocal, meta and hybrids) in combination with the spin disordered polymorphous description (SQS-PM)~\cite{PhysRevB.97.035107} we reveal that \emph{none} of the tested approaches is capable to concurrently deliver an accurate
description of the basic electronic and magnetic properties for both compounds. The main problem appears to be the proper treatment of magnetic itinerancy
and the relative stability of the PM and  $G$-AFM  ordering, a critical issue which is still debated experimentally~\cite{Shi2013, doi:10.7566/JPSCP.21.011013}.
Finally, by monitoring the changes of the ordered magnetic moment, band gap and volume across the transition
between NaOsO$_3$ and LiOsO$_3$ achieved by chemical doping (Na$\rightarrow$Li in NaOsO$_3$ or Li$\rightarrow$Na in LiOsO$_3$),
we demonstrated that it is the steric effect that controls the structural stability
($R3c$ vs. $Pnma$) and the gap opening (metallic vs. insulating state).

%--------------------------------------------------------------------------------
\section{Computational Details}\label{sec:details}
%--------------------------------------------------------------------------------

All first-principles calculations were performed by employing the projector augmented wave method~\cite{PhysRevB.50.17953} as implemented in the Vienna
\emph{Ab initio} Simulation Package (VASP)~\cite{PhysRevB.47.558, PhysRevB.54.11169} with the inclusion of the SOC.
 A plane-wave cutoff of 600 eV was used for both LiOsO$_3$ and NaOsO$_3$.
10$\times$10$\times$10 and  8$\times$6$\times$8 $\Gamma$-centered $k$-point grids generated by the Monkhorst-Pack scheme
were used for the rhombohedral LiOsO$_3$ unit cell and  orthorhombic NaOsO$_3$ unit cell,  respectively.
The ISOTROPY~\cite{ISOTROPY} and AMPLIMODES~\cite{Orobengoa:ks5225} programs were employed
to determine the group-subgroup relationships and perform the symmetry-adapted soft mode analysis.

In order to seek a common and consistent XC functional that can describe both compounds reasonably well, we have assessed the local density approximation (LDA)~\cite{PhysRevB.23.5048} in the parametrization of Ceperly and Alder~\cite{PhysRevLett.45.566},
the generalized gradient approximation (GGA) functional Perdew-Burke-Ernzerhof (PBE)~\cite{PhysRevLett.77.3865}
as well as its improved version for solid (PBEsol)~\cite{PhysRevLett.100.136406},
the strongly constrained appropriately normed  (SCAN) meta-GGA functional~\cite{PhysRevLett.115.036402}
and the hybrid functional HSE06~\cite{doi:10.1063/1.2404663}.
All calculations where done within the noncollinear DFT+$U$ framework~\cite{PhysRevMaterials.3.083802},
based on experimental low-temperature structural parameters of LiOsO$_3$~\cite{Shi2013} and  NaOsO$_3$~\cite{PhysRevB.80.161104}.
The conjugate gradient algorithm~\cite{RevModPhys.64.1045} was used for the electronic optimization with
an accuracy such that the total energy difference was less than $10^{-6}$ eV between iterations.
In order to quantify the strength of correlation effects, we computed $U$ from the cRPA
within the ``$t_\text{2g}/t_\text{2g}$" scheme~\cite{PhysRevB.86.165105} based on a nonmagnetic band structure.
For more details and notations about cRPA, we refer to Refs.~\cite{Merzuk2015, PhysRevMaterials.2.075003}. The matrix elements of
on-site Coulomb $U_{ij}$ and exchange $J_{ij}$ interactions calculated with LDA (without SOC) are given in Table~\ref{Tab:cRPA_matrix_t2g},
which yields an orbital-averaged  $U$ of 1.94 eV and 1.86 eV for LiOsO$_3$ and NaOsO$_3$, respectively.
Note that the calculated $U$ values are insensitive to the specific functional used (e.g., the difference is less than 0.05 eV between LDA and PBE).

\begin{table}%[h!]
\footnotesize
\caption {
On-site Coulomb $U_{ij}$ and exchange $J_{ij}$ interactions (in eV)
($i$ and $j$ represent $t_\text{2g}$ orbitals)
within the $t_\text{2g}/t_\text{2g}$ scheme for $R3c$-LiOsO$_3$ and NaOsO$_3$ using the LDA functional without SOC.}
\begin{ruledtabular}
\begin{tabular}{lcccccccc}
                                   & \multicolumn{3}{c}{$U_{ij}$}  & & \multicolumn{3}{c}{$J_{ij}$}  \\
                                   \hline
 LiOsO$_3$       & $d_\text{xz}$ & $d_\text{yz}$ & $d_\text{xy}$  && $d_\text{xz}$ & $d_\text{yz}$ & $d_\text{xy}$  \\
$d_\text{xz}$       & 2.33 & 1.74  & 1.74   && --      & 0.25 & 0.25  \\
$d_\text{yz}$       & 1.74 & 2.33 & 1.74   && 0.25 & --      & 0.25  \\
$d_\text{xy}$       & 1.74  & 1.74  & 2.33  && 0.25 & 0.25 & -- \\
\hline
NaOsO$_3$       & $d_\text{xz}$ & $d_\text{yz}$ & $d_\text{xy}$  && $d_\text{xz}$ & $d_\text{yz}$ & $d_\text{xy}$  \\
$d_\text{xz}$       & 2.22 & 1.66 & 1.67  && --      & 0.23 & 0.24  \\
$d_\text{yz}$       & 1.66 & 2.27 & 1.68  && 0.23 & --      & 0.24  \\
$d_\text{xy}$       & 1.67 & 1.68 & 2.27  && 0.24 & 0.24 & --
\end{tabular}
\end{ruledtabular}
\label{Tab:cRPA_matrix_t2g}
\end{table}

To clarify the geometric steric effect due to the different ionic radii of Li$^+$ and Na$^+$,
two computational experiments were designed by considering Na-doped $R3c$-LiOsO$_3$ and Li-doped $Pnma$-NaOsO$_3$.
The alloy structures were modeled using the special quasirandom structure (SQS) method~\cite{PhysRevLett.65.353}
as implemented in the ATAT package~\cite{VANDEWALLE2002539,VANDEWALLE201313}.
The SQS method mimics the disordered atomic configurations within a supercell of limited size
in terms of the correlation functions in the cluster expansion method.
For the Na-doped LiOsO$_3$, a supercell with 120 atoms was used,
whereas for the Li-doped NaOsO$_3$, a supercell with 80 atoms was employed.
A $4\times4\times4$ $k$-point grid was used to sampled the Brillouin zone (BZ) of all the supercells.
The supercells were fully relaxed (including the cell shape and atomic positions)
until the Hellmann-Feynman forces acting on each atom were less than 10 meV/$\AA$.

The phonon dispersions and density of states (DOS) were calculated
by finite displacements using the Phonopy code~\cite{PhysRevB.78.134106}.
For the cubic perovskite phases, a supercell with 135 atoms was used, whereas a $2\times2\times2$ supercell
was utilized for both $R3c$-LiOsO$_3$ (80 atoms) and  $Pnma$-NaOsO$_3$ (160 atoms).
For all phonon calculations, a $3\times3\times3$ $k$-point grid was used to sampled the BZ.
Test results show that the phonon DOS are converged with respect to the chosen supercell size and $k$-point grid.

%--------------------------------------------------------------------------------
\section{Results}\label{sec:results}
%--------------------------------------------------------------------------------

%--------------------------------------------------------------------------------
\subsection{Crystal structure and symmetry mode analysis}\label{sec_results:structures}
%--------------------------------------------------------------------------------

From a theoretical perspective, the different structural symmetries of NaOsO$_3$ ($Pnma$) and LiOsO$_3$ ($R3c$)
can be understood in terms of the different Goldschmidt tolerance factor
 $t=\frac{r_{\rm A}+r_{\rm O}}{\sqrt{2}(r_{\rm Os}+r_{\rm O})}$ ($r$ denotes the ionic radius and A=Li/Na).
In fact, the tolerance factor is often taken as an indicator for the degree of distortion of perovskites~\cite{Helmut2007,Benedek2013}:
$t$=1 represents the ideal conditions upon which the perovskite structure assumes its ideal cubic symmetry, which is generally stable
in the range $0.9<t<1$; $t>$1 favors a hexagonal structure, whereas $0.71<t<0.9$ yields
rhombohedral or orthorhombic structures~\cite{Schinzer2012}. With  a tolerance factor of 0.75 (LiOsO$_3$) and 0.84 (NaOsO$_3$),
these two Os-based perovskites are predicted to assume rhombohedral or orthorhombic phase, respectively~\cite{Schinzer2012},
since the slightly smaller $t$ of LiOsO$_3$ (originating from the smaller ionic radius of Li$^+$) should result in a more distorted rhombohedral structure.
To confirm these expectation we have conducted a symmetry analysis of the phonon dispersions of LiOsO$_3$ and NaOsO$_3$.

\begin{figure}%[h!]
\begin{center}
\includegraphics[width=0.5\textwidth,clip]{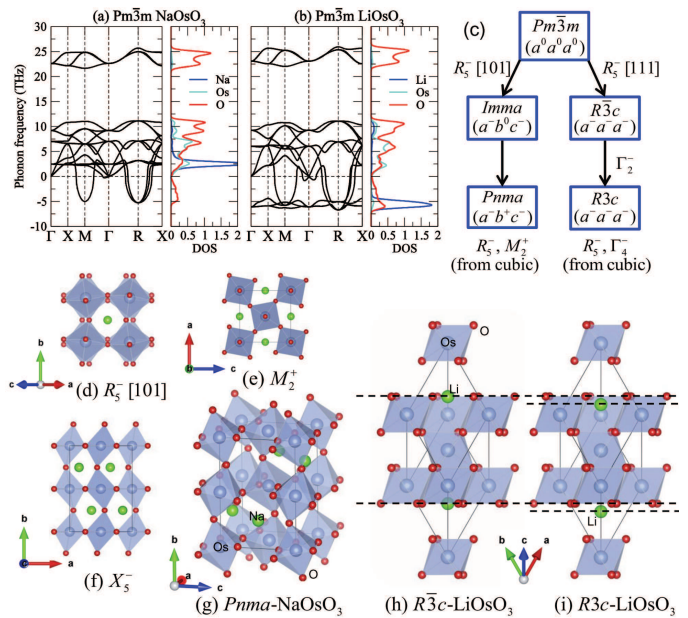}
\end{center}
 \caption{Comparison of LDA calculated phonon dispersions and partial DOS for cubic ($Pm\bar{3}m$)
 (a) NaOsO$_3$ and (b) LiOsO$_3$.    Imaginary frequencies are shown as negative values.
 (c) A diagram showing the group-subgroup relationships along with
 corresponding distortion modes, which are shown in (d) for $R^-_5$,
 (e) for $M^+_2$, and (f) for $X^-_5$ following the notations of Miller and Love~\cite{Miller1967}.
 For different space groups, the octahedral tilts/rotations represented in the Glazer notations~\cite{Glazer1972}  are also given.
  (g), (h), and (i) show the crystal structures of $Pnma$-NaOsO$_3$, $R\bar{3}c$-LiOsO$_3$, and $R3c$-LiOsO$_3$, respectively.
  The polar ferroelectric mode $\Gamma^-_2$ associated with the $R\bar{3}c$ to ferroelectric-like $R3c$ phase can be seen by comparing (i) to (h).
 Structural models were generated with VESTA~\cite{Momma:db5098}.
}
\label{fig:mode_analysis}
\end{figure}

Figs.~\ref{fig:mode_analysis}(a) and (b) show the calculated phonon dispersions and partial phonon DOS for cubic NaOsO$_3$ and LiOsO$_3$ using the LDA
functional, with optimized LDA lattice parameters. One can see that in general the two compounds show similar phonon dispersions except for the soft phonon modes.
For cubic LiOsO$_3$ the soft phonons are dominated by Li and O atoms, whereas for cubic NaOsO$_3$ the negative frequencies originate only from O atoms.
Both phonon dispersions share structural instability at the $R$ and $M$ points, but LiOsO$_3$ exhibits an additional soft mode at the zone center ($\Gamma$ point).
The $R$ mode corresponds to the antiphase octahedral rotation mode $R^-_5$ along the [101] and [111] axis
for NaOsO$_3$ [Fig.~\ref{fig:mode_analysis}(d)] and LiOsO$_3$, respectively.
By moving along this mode, the cubic phases of NaOsO$_3$ and LiOsO$_3$ reduce to the orthorhombic $Imma$ structure and
centrosymmetric rhombohedral $R\bar{3}c$ structure [Fig.~\ref{fig:mode_analysis}(h)], respectively.
Symmetry-adapted soft phonon analysis indicates that the instabilities at $M$ and  $\Gamma$
are due to the in-phase octahedral rotations [$M^+_2$, Fig.~\ref{fig:mode_analysis}(e)]
and the ferroelectric distortions associated with the displacements of Li atoms along the polar [111] axis
 [$\Gamma^-_2$, comparing Fig.~\ref{fig:mode_analysis}(i) to Fig.~\ref{fig:mode_analysis}(h)].
Further condensing these modes leads to the formation of $Pnma$-NaOsO$_3$ [Fig.~\ref{fig:mode_analysis}(g)]
and noncentrosymmetric $R3c$-LiOsO$_3$ [Fig.~\ref{fig:mode_analysis}(i)].
A diagram showing the group-subgroup relationships is shown in Fig.~\ref{fig:mode_analysis}(c).
It is worthy noting that an additional octahedral tilt mode $X^-_5$ occurring in
NaOsO$_3$ [Fig.~\ref{fig:mode_analysis}(f)] is a secondary mode
and it appears as a consequence of the combined effect of the primary modes $R^-_5$ and $M^+_2$.
With respect to the parent cubic phase, the normalized amplitudes of the modes $R^-_5$,  $M^+_2$, and $X^-_5$ for
the experimental $Pnma$-NaOsO$_3$ structure are estimated to be about 0.72 $\AA$, 0.50 $\AA$, and 0.28 $\AA$, respectively,
while those associated with the modes $R^-_5$ and $\Gamma^-_4$ for the rhombohedral $R3c$-LiOsO$_3$ structure are about 1.22 $\AA$ and 0.47 $\AA$, respectively.
Our mode analysis on LiOsO$_3$ is consistent with Ref.~\cite{PhysRevB.89.201107}.

%--------------------------------------------------------------------------------
\subsection{Assessing the XC functionals on electronic and magnetic properties}\label{sec_results:functionals}
%--------------------------------------------------------------------------------

\begin{figure}%[h!]
\begin{center}
\includegraphics[width=0.5\textwidth,clip]{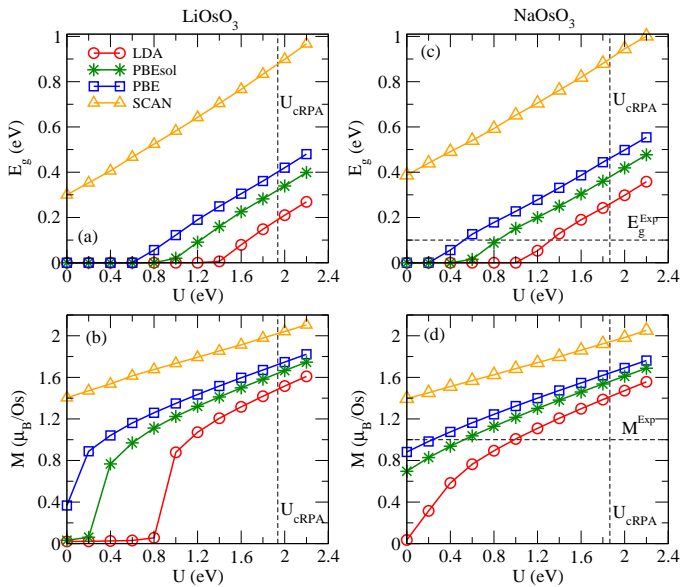}
\end{center}
 \caption{Comparison of magnetic moments $M$ ($\mu_{\rm B}$/Os) and band gaps $E_{\rm g}$ (eV)
 for LiOsO$_3$ [(a) and (b)] and NaOsO$_3$ [(c) and (d)] calculated by the DFT+$U$+SOC approach as a function of $U$
 using different XC functionals:  LDA (circles), PBE (squares), PBEsol (starts), and SCAN (triangles).
 Note that here the experimental structures are used for all calculations.
The cRPA calculated orbital-averaged $U$ values for both systems  and experimental magnetic moments and band gaps for
NaOsO$_3$ are indicated as dashed lines.
}
\label{fig:compare_functionals}
\end{figure}

It is well known that the specific form of the XC functional plays an important role in first-principles DFT simulations
and that finding an XC functional capable to account for the basic ground state properties is nontrivial, especially for complex materials.
LiOsO$_3$ and NaOsO$_3$ represent typical examples that pose great challenges for the choice of an XC functional
and no consensus has been achieved yet in literature due to a lack of proper scrutinization.
For instance, NaOsO$_3$ was studied using LDA in Refs.~\cite{PhysRevB.80.161104,PhysRevB.85.174424,PhysRevB.87.115119},
and using PBE in Ref.~\cite{PhysRevB.94.241113}. Similarly, for LiOsO$_3$, LDA was used in Refs.~\cite{PhysRevB.90.195113, PhysRevB.90.094108,doi:10.1002/pssr.201800396},
while PBE was employed in Refs.~\cite{PhysRevB.89.201107,PhysRevB.91.064104} and PBEsol in Ref.~\cite{Paredes2018}.
Considering the quantitative and --to some extent-- qualitative discrepancies between the results obtained by different XC functionals
and in order to achieve a trustable and convincing comparative study between these two systems,
we have performed a systematic assessment of the performance of LDA, PBE, PBEsol, and SCAN within a DFT+$U$+SOC framework
(with $U$ ranging from 0 to 2.4~eV) as well as HSE06 for the prediction of band gaps and magnetic properties of NaOsO$_3$ and LiOsO$_3$.
The GS magnetic states are determined by comparing the total energy difference
between the two energetically favorable configurations, $G$-type AFM state and nonmagnetic state, for each $U$ value~\cite{doi:10.1002/pssr.201800396}.
The results displayed in Fig.~\ref{fig:compare_functionals} show that \emph{none} of the considered functionals is capable to
simultaneously predict an insulating magnetic state for NaOsO$_3$ and a non-magnetic metallic state for LiOsO$_3$.

As a general and expected trend, we remark that the inclusion on the onsite $U$ tends to favor an insulating solution
and to establish a magnetic ordering: Both band gap and local magnetic moment increase with increasing $U$,
but a single value of $U$ cannot establish the desired ground states in both systems.
The situation is particularly problematic for LiOsO$_3$, as discussed in more details in the following.

First, we note that SCAN, typically considered to be a rather accurate scheme, tends to overestimate the magnetic moments for both
compounds [Figs.~\ref{fig:compare_functionals}(b) and (d)], as it does for itinerant electron ferromagnets~\cite{PhysRevB.98.094413, PhysRevLett.121.207201}.
Also, it overestimates the band gap for NaOsO$_3$ [Figs.~\ref{fig:compare_functionals}(c)] and wrongly predicts a magnetic insulating state for LiOsO$_3$ [Figs.~\ref{fig:compare_functionals}(a) and (b)]. We note that similarly to SCAN, HSE06+SOC with default screening length ($\mu$=0.2)
and exact exchange mixing parameters ($\alpha$=0.25) delivers an even lager band gap and magnetic moment:
It gives an incorrect magnetic insulating state for LiOsO$_3$ with a gap of 1.25 eV and a moment of 1.70 $\mu_B$/Os,
while for NaOsO$_3$ it predicts a gap of 1.43 eV and a moment of 1.67 $\mu_B$/Os.
It should be noted, however, that hybrid functionals are sensitive to the choice of $\mu$ and $\alpha$ parameters
and for moderately correlated itinerant systems the optimal value should deviate substantially from the default ones~\cite{PhysRevB.86.235117,Liu_2019}.
In the following we will focus on a detailed discussion of the LDA, PBE, and PBEsol results.

\begin{table}%[t!]
\caption {
The critical $U_c$ (in eV) required for the magnetically driven MIT and the corresponding critical magnetic moment $M_c$ (in $\mu_B$/Os) calculated at $U_c$
for different XC functionals. Since the SCAN alone already opens the band gap, a negative $U_c$ is obtained for the onset of the MIT.}
\begin{ruledtabular}
\begin{tabular}{lrcrc}
& \multicolumn{2}{c}{LiOsO$_3$}  & \multicolumn{2}{c}{NaOsO$_3$}  \\
\cline{2-3} \cline{4-5}
   & $U_c$ & $M_c$ & $U_c$ & $M_c$  \\
\hline
LDA    &   1.4 &  1.18 &     1.0 &  1.01 \\
PBEsol &   0.8 &  1.11 &     0.4 &  0.94 \\
PBE    &   0.6 &  1.16 &     0.2 &  0.98 \\
SCAN   &$-$1.0 &  1.07 &  $-$1.4 &  0.99 \\
\end{tabular}
\end{ruledtabular}
\label{Tab:Uc_moment}
\end{table}

In line with previous studies~\cite{PhysRevB.80.161104,PhysRevB.85.174424,PhysRevB.94.241113,PhysRevB.87.115119},
without $U$, LDA, PBE, or PBEsol fail to open the band gap in NaOsO$_3$ [Fig.~\ref{fig:compare_functionals}(b)], and while PBE and PBEsol
find a sizable local moment, LDA favors a nonmagnetic solution, in disagreement with experimental observations.
The situation in LiOsO$_3$ is similar with the only exception that PBEsol does not stabilize any magnetic solution,
in this case in line with the experimentally observed metallic nonmagnetic ground state.

With increasing $U$, the situation remains problematic. Above a certain critical $U_c$ both systems undergo a MIT
and both the gap and magnetic moments grow almost linearly as a function of $U$.  The values of $U_c$  are listed in Table~\ref{Tab:Uc_moment}.
A positive outcome of the calculations is that $U_c$ is systematically lower in NaOsO$_3$ than in LiOsO$_3$
implying that there exists a $U$ window for each functional where NaOsO$_3$ is insulating and LiOsO$_3$ metallic (LDA: 1.0-1.4 eV; PBEsol: 0.4-0.8 eV; PBE: 0.2-0.4 eV).
The downside is that within these $U$ ranges, \emph{both} systems are found to be magnetic, which is good for NaOsO$_3$ but in apparent disagreement with experiment for LiOsO$_3$.
More precisely, PBE+$U$+SOC yields an ordered magnetic moment for LiOsO$_3$ for all $U$ values.
On the other hand, LDA and PBEsol yield a nonmagnetic solution in the low-$U$ limit, but a magnetic moment develops for $U$ values larger than $U_c$,
 which is therefore outside the $U$ range indicated above. As soon as the $U$ reaches $U_c$ a well established magnetic moment $M_c$ of about 1 $\mu_B$
is found for all functionals as reported in Table~\ref{Tab:Uc_moment}.

Summing up, LiOsO$_3$ is nonmagnetic and metallic for $U\leq0.2$ eV (PBEsol) and  $U\leq0.8$ eV (LDA).
In this $U$ range, however, NaOsO$_3$ is magnetic but always metallic. The most likely cause of this apparent disagreement
is the shortcoming of mean-field DFT in the LDA or GGA to account for magnetic fluctuations in itinerant magnets~\cite{PhysRevB.86.064437,PhysRevB.73.205121},
and this leads to a systematic overestimation of the local ordered magnetic moment.

It needs to be noted that although at low temperature a long-range magnetic order is absent in LiOsO$_3$~\cite{Shi2013},
the Curie-Weiss-like behaviour observed below $T_s$~\cite{Shi2013} and the $\mu$SR experiments~\cite{doi:10.7566/JPSCP.21.011013}
suggest a disordered PM ground state. Within DFT it is challenging to model the PM state.
Recently, G. Trimarchi \emph{et al.}~\cite{PhysRevB.97.035107} proposed a polymorphous description for the spin disordered state,
which is realized by a supercell calculation modeled in the SQS manner (termed SQS-PM)~\cite{PhysRevB.97.035107}.
Using this method, the gap opening and orbital ordering of the paramagnetic phases
of the transition-metal monoxides~\cite{PhysRevB.97.035107} as well as the 3$d$ perovskite oxides~\cite{Varignon2019}
are reasonably well described.

\begin{figure}%[h!]
\begin{center}
\includegraphics[width=0.49\textwidth,clip]{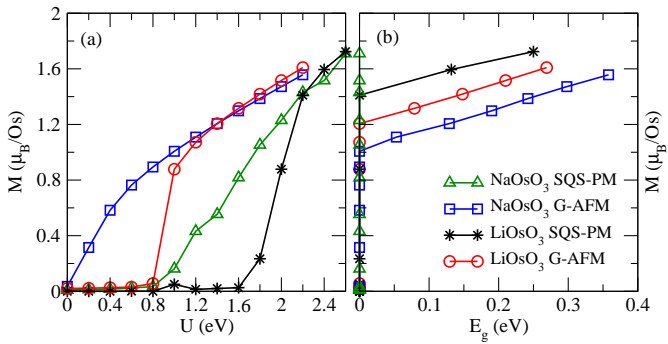}
\end{center}
 \caption{ LDA+$U$+SOC calculated magnetic moments $M$ ($\mu_{\rm B}$/Os) and band gaps $E_{\rm g}$ (eV)
as a function of $U$  for LiOsO$_3$ and NaOsO$_3$ with $G$-AFM and SQS-PM polymorphous descriptions.
}
\label{fig:compare_LDA}
\end{figure}

Aiming to improve the description of the PM phase in LiOsO$_3$, we have applied this SQS-PM approach using the LDA functional.
The results are shown in Fig.~\ref{fig:compare_LDA} where we show the correlation between $U$ and the magnetic moment $M$ [Fig.~\ref{fig:compare_LDA}(a)]
and the correlation between $M$ and the band gap $E_g$ [Fig.~\ref{fig:compare_LDA}(b)].
One can observe that for $U$ smaller than 1.6 eV, the SQS-PM (stars) predicts a nonmagnetic state.
As $U$ increases, the magnitude of the disordered magnetic moment increases and a MIT appears for $U$ larger than 2 eV.
By contrast, the SQS-PM description of NaOsO$_3$ always gives metallic solutions for the considered $U$ values, in line with the observation that only
$G$-AFM is capable to open the gap, whereas all other magnetic orderings yield a metallic solutions~\cite{PhysRevB.85.174424}.
Although the larger critical $U_c$ required for the MIT in the SQS-PM phase seems to mitigate the above-mentioned
issues of the XC functionals, the SQS-PM solutions of LiOsO$_3$ are always higher in energy than the $G$-AFM ordered phases.
Taking $U$=1.8 eV for instance, the energy difference is about 71 meV/f.u.
Therefore, even the SQS-PM method is not a satisfactory solution and one might
have to resort to either a new XC functional or a new method to treat itinerant
magnetism in DFT~\cite{PhysRevB.86.064437,PhysRevB.96.035141,Sharma2018}.

\begin{figure}%[t!]
\begin{center}
\includegraphics[width=0.49\textwidth,clip]{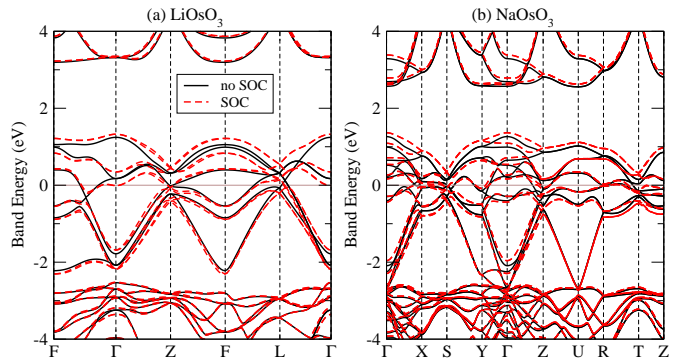}
\end{center}
 \caption{Comparison of band structures
 for nonmagnetic (a) $R3c$-LiOsO$_3$ and (b) NaOsO$_3$
 calculated by LDA (black lines) and LDA+SOC (red dashed lines).
}
\label{fig:bandsU0}
\end{figure}

We conclude this section with a remark on the correlation strength of these compounds.
From Fig.~\ref{fig:compare_functionals} one can see that the cRPA estimated $U$ values (dashed lines) are very large
and fall in a range in which both systems are magnetic insulator. We have previously reported that SOC effects could induce a considerable
renormalization of the Coulomb interaction in NaOsO$_3$ of about 1 eV, placing NaOsO$_3$ in the moderately correlated regime.
This spin-orbit renormalization was also found to be necessary to correctly describe the Lifshitz transition of NaOsO$_3$~\cite{PhysRevB.94.241113}.
The comparison of the band structures with and without SOC for nonmagnetic LiOsO$_3$ and NaOsO$_3$ (Fig.~\ref{fig:bandsU0})
indeed shows that the inclusion of SOC increases the bandwidth of the $t_{\rm 2g}$ states and thus enhances electron mobility.
This reduces the correlation strength and leads to smaller $U$ value as compared to the one obtained without SOC~\cite{PhysRevB.94.241113,PhysRevB.98.205128}.
However, a precise quantification of the SOC renormalization effect is a difficult task requiring the inclusion of SOC in the cRPA calculation.
Unfortunately, to our knowledge, no cRPA implementation is available to compute $U$ with SOC because of technical complexity
in treating the complex-valued Wannier spinors within the cRPA scheme.
cRPA calculations without SOC suggest that the two compounds have very similar Coulomb parameters (see Table~\ref{Tab:cRPA_matrix_t2g}),
and therefore, in the following calculations we will adopt the same $U$ for both materials.

%--------------------------------------------------------------------------------
\subsection{Cooperative steric-magnetic driven MIT}\label{sec_results:SQS}
%--------------------------------------------------------------------------------

Since GGA overestimates the magnetic moment for itinerant magnets even more than LDA~\cite{PhysRevB.86.064437}
and LDA performs generally better than GGA in predicting ferroelectric properties~\cite{PhysRevB.96.035143},
in the following calculations the LDA with $U$ = 1.2 eV is employed [results obtained for $U$=1.4 eV show very similar trends (not shown)].

\begin{figure}%[h!]
\begin{center}
\includegraphics[width=0.45\textwidth,clip]{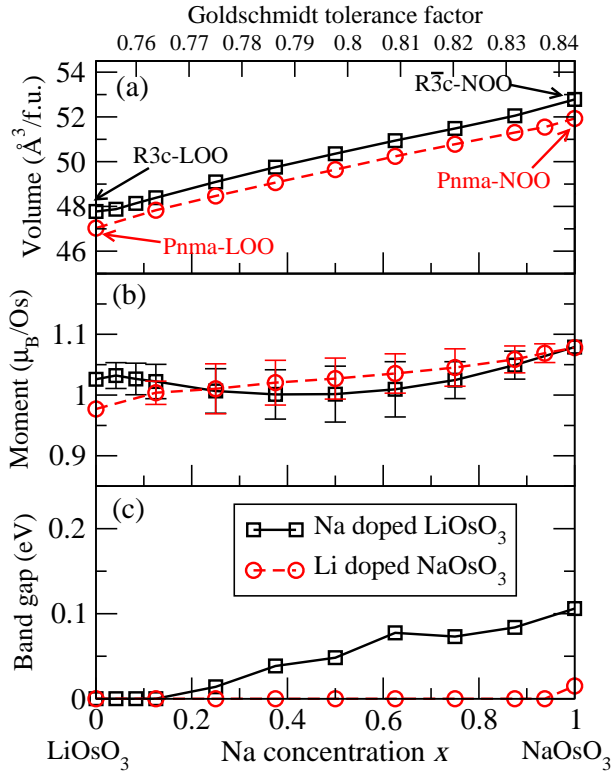}
\end{center}
 \caption{LDA+$U$+SOC  ($U$=1.2 eV) calculated (a) volumes, (b) Os sites averaged magnetic moments, and (c) band gaps of
 Na-doped LiOsO$_3$ (LOO) (squares)  and Li-doped NaOsO$_3$ (NOO) (circles)
 as a function of Na concentration $x$ calculated by SQS supercell calculations. The standard deviation of
 the fluctuating magnetic moments arising from the disorder effects is shown as error bars.
}
\label{fig:SQS_LDApUSOC}
\end{figure}

As discussed previously, it is the difference in the tolerance factor that dictates the different crystal structures in LiOsO$_3$ and NaOsO$_3$.
Then a natural question is whether this is also the origin of their distinct electronic properties.
To verify this hypothesis, we designed two computational experiments to track the transition between LiOsO$_3$ and
NaOsO$_3$ via chemical doping using  LDA+$U$+SOC in combination with SQS. We have inspected the following two scenarios.
($i$) Doping $R3c$-LiOsO$_3$ with Na. In this way, we study how Na doping affects the metallic ground state of LiOsO$_3$.
($ii$) Doping $Pnma$-NaOsO$_3$ with Li, where we control the influence of Li doping on the metallic state of NaOsO$_3$.
For different Na (or Li) contents we have computed the optimized volume, the magnetic moment and the band gap in the $R3c$ and $Pnma$ phase.
The results are shown in Fig.~\ref{fig:SQS_LDApUSOC}.

Let us first focus on Na-doped LiOsO$_3$ (squares in Fig.~\ref{fig:SQS_LDApUSOC}).
As expected, the volume (tolerance factor) increases  almost linearly
as the Na doping concentration $x$ increases. At $x\approx$20\% a MIT appears
and the band gap increases further as $x$ increases. The magnetic moment increases very slowly
within the standard deviation of the fluctuating moments induced by disorder effects.
Analogously, in Li-doped NaOsO$_3$ (circles in Fig.~\ref{fig:SQS_LDApUSOC})
as the Li concentration (1-$x$) increases, the volume and the tolerance factor
decrease and the insulator-to-metal transition occurs at a low Li concentration of 6.25\%.

\begin{figure}%[h!]
\begin{center}
\includegraphics[width=0.45\textwidth,clip]{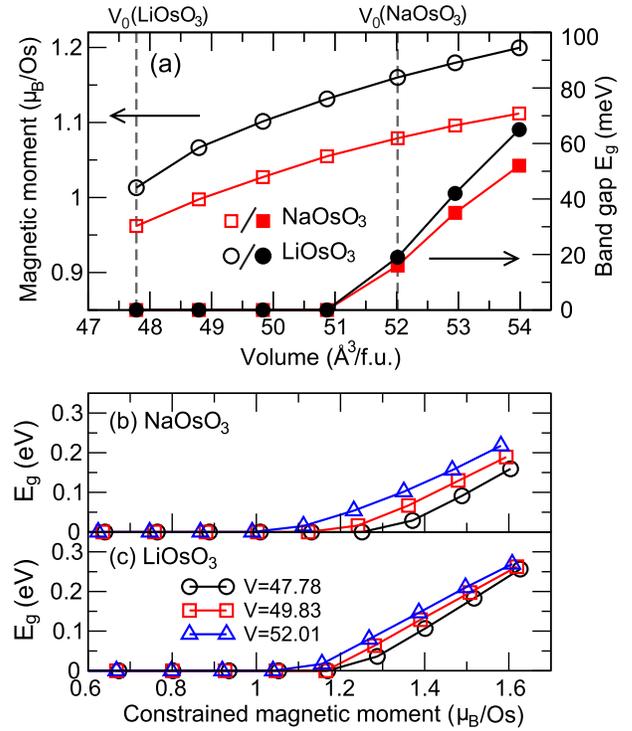}
\end{center}
 \caption{(a)  LDA+$U$+SOC ($U$=1.2 eV) calculated band gaps $E_g$
 and magnetic moments as a function of the system volume for $Pnma$-NaOsO$_3$ (squares) and $R3c$-LiOsO$_3$ (circles).
 The optimized volumes at ambient pressure are indicated.
 (b) and (c) show the calculated band gap as a function of the constrained magnetic moment for three fixed volumes
 of NaOsO$_3$ and LiOsO$_3$, respectively.
}
\label{fig:gap_moment_vs_volume}
\end{figure}

\begin{figure*}%[h!]
\begin{center}
\includegraphics[width=1.00\textwidth,clip]{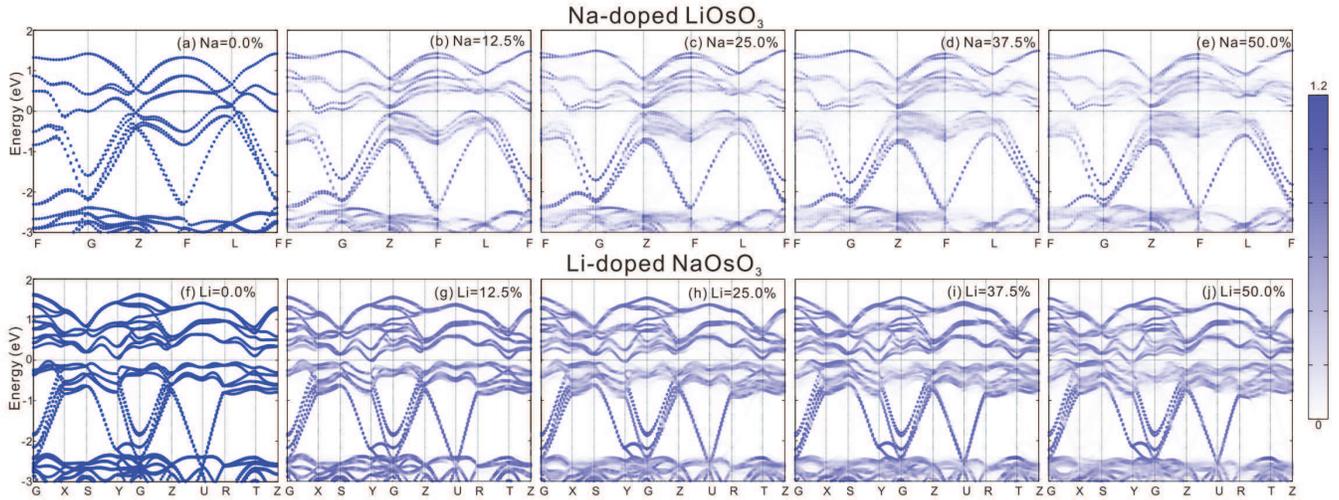}
\end{center}
 \caption{Evolution of the effective band structure (EBS) in Na-doped LiOsO$_3$ (upper panels)  and Li-doped NaOsO$_3$  (bottom panels)  (selected concentrations),
unfolded in the corresponding primitive cell by means of the band unfolding technique~\cite{PhysRevLett.104.236403,PhysRevB.94.195145}.
The lateral bar indicates the amount of the Bloch character. The sharpness of the EBS reflects the effect of the chemical disorder.
}
\label{fig:bands}
\end{figure*}

The overall similar trends in Na-doped LiOsO$_3$ and Li-doped NaOsO$_3$
convey a clear conclusion: The electronic ground state is mainly controlled by steric effects.
The larger atomic radius of Na increases the volume and thus favors the onset of the insulating state.
By replacing back Na with Li in the insulating phase of Na-doped LiOsO$_3$
and performing electronic self-consistent calculations while keeping the atom's positions fixed at the corresponding Li sites,
it is found that the band gap remains open.
Expectedly, by fixing the volume, $R3c$-LiOsO$_3$ displays a larger
magnetic moment and a larger tendency to become insulating than $Pnma$-NaOsO$_3$, as shown in Fig.~\ref{fig:gap_moment_vs_volume}(a).
It is also obvious that it is  the larger/smaller ground-state volume of NaOsO$_3$/LiOsO$_3$ that makes the system insulating/metallic.
In addition, it is found that the band structure at a fixed crystal structure is essentially insensitive to the Na/Li cation (not shown).
All these facts imply that the origin of the different electronic (metallic vs. insulating)
properties of the two compounds is primarily driven by steric effects.
The detailed changes on the effective band structure due to doping are displayed in Fig.~\ref{fig:bands},
highlighting the emergence of the MIT in Na-doped LiOsO$_3$ and Li-doped NaOsO$_3$.

However, it is worth noting that the presence of magnetic order also plays an important role in
the onset of the MIT. As shown in Figs.~\ref{fig:gap_moment_vs_volume}(b) and (c),
the MIT appears \emph{only} when the magnetic moment is larger than a critical value, which
gets progressively reduced by increasing the volume.
The fact that a larger volume favors a larger magnetic moment
and the larger magnetic moment in turn assists the opening of the band gap
suggests that the onset of MIT for the two compounds is driven by
a \emph{cooperative} steric and magnetic effect.

%--------------------------------------------------------------------------------
\subsection{Structural stability and phase transition}\label{sec_results:phonons}
%--------------------------------------------------------------------------------

\begin{figure}%[h!]
\begin{center}
\includegraphics[width=0.48\textwidth,clip]{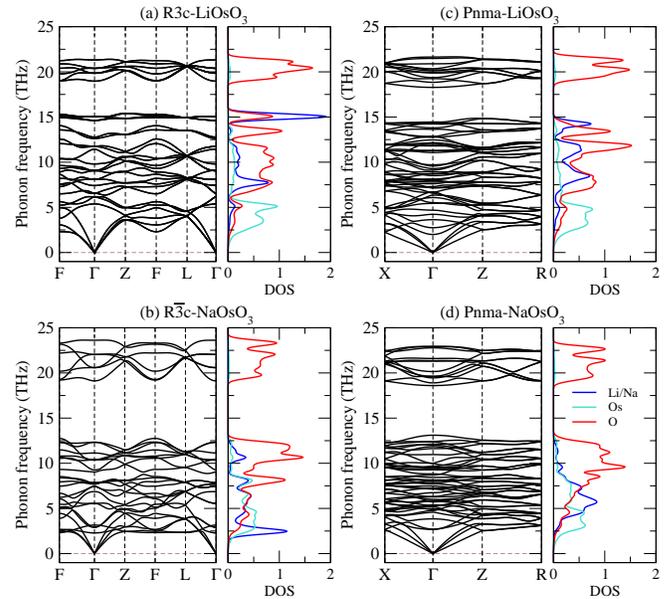}
\end{center}
 \caption{Comparison of LDA+$U$+SOC ($U$=1.2 eV) calculated phonon dispersions
 and partial DOS for LiOsO$_3$ with (a) $R3c$ and (c) $Pnma$ symmetries
 and for NaOsO$_3$ with (b) $R\bar{3}c$ and (d) $Pnma$ symmetries at zero pressure.
 The results  obtained from $U$=1.4 eV are similar.
}
\label{fig:phonon_compare_all}
\end{figure}

\begin{figure}%[h!]
\begin{center}
\includegraphics[width=0.48\textwidth,clip]{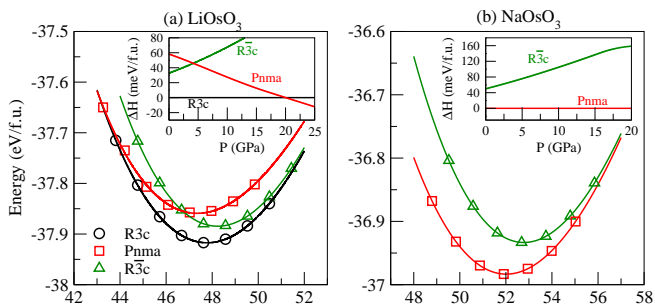}
\end{center}
 \caption{LDA+$U$+SOC ($U$=1.2 eV) calculated total energies as a function of the
  volume for (a) LiOsO$_3$ and (b) NaOsO$_3$ with different symmetries.
  The solid lines are obtained by fitting with the Birch-Murnaghan equation of state~\cite{PhysRev.71.809}.
  The insets show the zero-temperature enthalpy difference $\Delta H$ as a function of pressure.
  The $R3c$-LiOsO$_3$ and $Pnma$-NaOsO$_3$ phases are taken as references.
}
\label{fig:E_V_all}
\end{figure}

Now we turn to discussing dynamical properties and possible structural phase transitions in LiOsO$_3$ and NaOsO$_3$.
As expected, the GS phases of  $R3c$-LiOsO$_3$ and $Pnma$-NaOsO$_3$ are dynamically stable,
as revealed by the phonon dispersions displayed in Figs.~\ref{fig:phonon_compare_all}(a) and (d).
Although the $Pnma$-LiOsO$_3$ and $R\bar{3}c$-NaOsO$_3$ phases at ambient pressure
are energetically less favorable than the corresponding GS phases by $<$60 meV/f.u. (see Fig.~\ref{fig:E_V_all}),
they turn out to be dynamically stable, since no soft mode appears in the vibrational spectra  [see Figs.~\ref{fig:phonon_compare_all}(b) and (c)].
We also mention that the NaOsO$_3$ with polar $R3c$ symmetry is unstable and reduces to the nonpolar $R\bar{3}c$ symmetry
after structural relaxations, suggesting that ferroelectric instabilities are not expected in NaOsO$_3$ under these conditions.
This is consistent with the phonon calculations for cubic NaOsO$_3$,
where no instability at the $\Gamma$ point is observed [Fig.~\ref{fig:mode_analysis}(a)].
For LiOsO$_3$, on the other hand, we predict a structural phase transition from the $R3c$ phase to the $Pnma$ phase
at a pressure of about 20 GPa [see the inset of Fig.~\ref{fig:E_V_all}(a)],
consistent with the theoretical findings of E. Aulestia \emph{et al.}~\cite{Paredes2018}.
Even above the transition pressure, we find that the $Pnma$-LiOsO$_3$ phase is dynamically stable (not shown),
indicating that such a phase transition can be achievable for LiOsO$_3$ in high-pressure experiments.
However, for NaOsO$_3$ our calculations do not discern any symmetry change within the considered pressure range:
The $Pnma$ phase is always stable and becomes progressively stabilized over the $R\bar{3}c$ phase by increasing pressure [Fig.~\ref{fig:E_V_all}(b)].

Interestingly, we also find that the $Pnma$-LiOsO$_3$ or $R\bar{3}c$-NaOsO$_3$ phases seem to be dynamically stable
only if the magnetic order is present. For instance, with a smaller $U$=0.8 eV, both $Pnma$-LiOsO$_3$ and $R\bar{3}c$-NaOsO$_3$
are found to be magnetically-ordered metals, without any soft phonons. If the magnetic moment is removed, e.g., by performing
a non-spin-polarized LDA calculation, soft phonons appear and the system becomes dynamically unstable, a further indication of the strong spin-lattice
effects in this class of compounds~\cite{Calder2015}.

%--------------------------------------------------------------------------------
\section{Conclusions}\label{sec:conlcusions}
%--------------------------------------------------------------------------------

In conclusion, by comparative \emph{ab initio} DFT+$U$+SOC calculations, we have systematically studied the strikingly distinct
structural, electronic, magnetic, and dynamical properties of the two chemically similar osmates perovskites, LiOsO$_3$ and NaOsO$_3$.

First, we find that none of the considered XC functionals (LDA, PBE, PBEsol, SCAN and HSE06)
is capable to accurately predict the correct electronic and magnetic ground state for both compounds simultaneously.
This drawback is mostly due to the difficulties of DFT (within local, semilocal and nonlocal treatment of XC effects) in treating the magnetic
fluctuations associated with the itinerant nature of LiOsO$_3$ and NaOsO$_3$. Neglecting fluctuations ultimately leads to an overestimation of magnetic moments.
The SQS-PM approach allows for an improved description in stabilizing a PM state in LiOsO$_3$,
but it incorrectly predicts a higher energy for the PM phase than for the $G$-AFM phase in LiOsO$_3$.
In comparison to NaOsO$_3$, LiOsO$_3$ is less magnetic due to its smaller volume and to induce the MIT it requires a larger critical $U_c$.
Though using a different $U$ in the two systems (for instance U$\approx$1.1~eV in NaOsO$_3$ and U$\approx$0.7~eV in LiOsO$_3$) would lead to a reasonable
description of the two distinct ground states, the verification of this hypothesis would require the \emph{ab initio} calculation of $U$ including SOC effects,
which is, however, currently not possible.

Second, by following the transition from one compound to the other via chemical doping, we clarify that it is the cooperative steric and magnetic effect that
controls the  electronic properties and drives the formation of the distinct metallic/insulating state in the two systems:
The larger/smaller volume of NaOsO$_3$/LiOsO$_3$ leads to a larger/smaller magnetic moment, which in turn assists the opening/closing of the band gap.

Finally, the different GS crystal structures ($R3c$ vs. $Pnma$) of LiOsO$_3$ and NaOsO$_3$
can be explained by purely steric effects and arise from the different Goldschmidt tolerance factors (0.75 vs. 0.84).
Moreover, we show that the energetically unfavorable phases of $Pnma$-LiOsO$_3$ and
$R\bar{3}c$-NaOsO$_3$ at ambient pressure are dynamically stable.
A pressure-induced structural phase transition from $R3c$ to $Pnma$ for LiOsO$_3$ is predicted,
whereas for NaOsO$_3$ the $Pnma$ phase is stabilized over the $R\bar{3}c$ phase by increasing pressure, suggesting that under these conditions
NaOsO$_3$, unlike LiOsO$_3$, does not seem to be prone to ferroelectric instabilities.

 \begin{acknowledgments}
 Useful discussions with Danilo Puggioni are gratefully acknowledged.
This work was supported by the Austrian Science Fund (FWF) within the SFB ViCoM (Grant No. F 41).
BK acknowledges support by NRF Grand No. 2018R1D1A1A02086051 and Max-Planck POSTECH/KOREA Research Initiative (No. 2016K1A4A4A01922028).
Supercomputing time on the Vienna Scientific cluster (VSC) is acknowledged.
\end{acknowledgments}

\bibliographystyle{apsrev4-1}
\bibliography{reference} %here in windows there is no surfix .bib

%merlin.mbs apsrev4-1.bst 2010-07-25 4.21a (PWD, AO, DPC) hacked
%Control: key (0)
%Control: author (72) initials jnrlst
%Control: editor formatted (1) identically to author
%Control: production of article title (-1) disabled
%Control: page (0) single
%Control: year (1) truncated
%Control: production of eprint (0) enabled
\begin{thebibliography}{70}%
\makeatletter
\providecommand \@ifxundefined [1]{%
 \@ifx{#1\undefined}
}%
\providecommand \@ifnum [1]{%
 \ifnum #1\expandafter \@firstoftwo
 \else \expandafter \@secondoftwo
 \fi
}%
\providecommand \@ifx [1]{%
 \ifx #1\expandafter \@firstoftwo
 \else \expandafter \@secondoftwo
 \fi
}%
\providecommand \natexlab [1]{#1}%
\providecommand \enquote  [1]{``#1''}%
\providecommand \bibnamefont  [1]{#1}%
\providecommand \bibfnamefont [1]{#1}%
\providecommand \citenamefont [1]{#1}%
\providecommand \href@noop [0]{\@secondoftwo}%
\providecommand \href [0]{\begingroup \@sanitize@url \@href}%
\providecommand \@href[1]{\@@startlink{#1}\@@href}%
\providecommand \@@href[1]{\endgroup#1\@@endlink}%
\providecommand \@sanitize@url [0]{\catcode `\\12\catcode `\$12\catcode
  `\&12\catcode `\#12\catcode `\^12\catcode `\_12\catcode `\%12\relax}%
\providecommand \@@startlink[1]{}%
\providecommand \@@endlink[0]{}%
\providecommand \url  [0]{\begingroup\@sanitize@url \@url }%
\providecommand \@url [1]{\endgroup\@href {#1}{\urlprefix }}%
\providecommand \urlprefix  [0]{URL }%
\providecommand \Eprint [0]{\href }%
\providecommand \doibase [0]{http://dx.doi.org/}%
\providecommand \selectlanguage [0]{\@gobble}%
\providecommand \bibinfo  [0]{\@secondoftwo}%
\providecommand \bibfield  [0]{\@secondoftwo}%
\providecommand \translation [1]{[#1]}%
\providecommand \BibitemOpen [0]{}%
\providecommand \bibitemStop [0]{}%
\providecommand \bibitemNoStop [0]{.\EOS\space}%
\providecommand \EOS [0]{\spacefactor3000\relax}%
\providecommand \BibitemShut  [1]{\csname bibitem#1\endcsname}%
\let\auto@bib@innerbib\@empty
%</preamble>
\bibitem [{\citenamefont {Witczak-Krempa}\ \emph {et~al.}(2014)\citenamefont
  {Witczak-Krempa}, \citenamefont {Chen}, \citenamefont {Kim},\ and\
  \citenamefont {Balents}}]{doi:10.1146/annurev-conmatphys-020911-125138}%
  \BibitemOpen
  \bibfield  {author} {\bibinfo {author} {\bibfnamefont {W.}~\bibnamefont
  {Witczak-Krempa}}, \bibinfo {author} {\bibfnamefont {G.}~\bibnamefont
  {Chen}}, \bibinfo {author} {\bibfnamefont {Y.~B.}\ \bibnamefont {Kim}}, \
  and\ \bibinfo {author} {\bibfnamefont {L.}~\bibnamefont {Balents}},\ }\href
  {\doibase 10.1146/annurev-conmatphys-020911-125138} {\bibfield  {journal}
  {\bibinfo  {journal} {Annual Review of Condensed Matter Physics}\ }\textbf
  {\bibinfo {volume} {5}},\ \bibinfo {pages} {57} (\bibinfo {year}
  {2014})}\BibitemShut {NoStop}%
\bibitem [{\citenamefont {Martins}\ \emph {et~al.}(2017)\citenamefont
  {Martins}, \citenamefont {Aichhorn},\ and\ \citenamefont
  {Biermann}}]{Martins_2017}%
  \BibitemOpen
  \bibfield  {author} {\bibinfo {author} {\bibfnamefont {C.}~\bibnamefont
  {Martins}}, \bibinfo {author} {\bibfnamefont {M.}~\bibnamefont {Aichhorn}}, \
  and\ \bibinfo {author} {\bibfnamefont {S.}~\bibnamefont {Biermann}},\ }\href
  {\doibase 10.1088/1361-648x/aa648f} {\bibfield  {journal} {\bibinfo
  {journal} {Journal of Physics: Condensed Matter}\ }\textbf {\bibinfo {volume}
  {29}},\ \bibinfo {pages} {263001} (\bibinfo {year} {2017})}\BibitemShut
  {NoStop}%
\bibitem [{\citenamefont {Erg\"onenc}\ \emph {et~al.}(2018)\citenamefont
  {Erg\"onenc}, \citenamefont {Kim}, \citenamefont {Liu}, \citenamefont
  {Kresse},\ and\ \citenamefont {Franchini}}]{PhysRevMaterials.2.024601}%
  \BibitemOpen
  \bibfield  {author} {\bibinfo {author} {\bibfnamefont {Z.}~\bibnamefont
  {Erg\"onenc}}, \bibinfo {author} {\bibfnamefont {B.}~\bibnamefont {Kim}},
  \bibinfo {author} {\bibfnamefont {P.}~\bibnamefont {Liu}}, \bibinfo {author}
  {\bibfnamefont {G.}~\bibnamefont {Kresse}}, \ and\ \bibinfo {author}
  {\bibfnamefont {C.}~\bibnamefont {Franchini}},\ }\href {\doibase
  10.1103/PhysRevMaterials.2.024601} {\bibfield  {journal} {\bibinfo  {journal}
  {Phys. Rev. Materials}\ }\textbf {\bibinfo {volume} {2}},\ \bibinfo {pages}
  {024601} (\bibinfo {year} {2018})}\BibitemShut {NoStop}%
\bibitem [{\citenamefont {Kim}\ \emph {et~al.}(2008)\citenamefont {Kim},
  \citenamefont {Jin}, \citenamefont {Moon}, \citenamefont {Kim}, \citenamefont
  {Park}, \citenamefont {Leem}, \citenamefont {Yu}, \citenamefont {Noh},
  \citenamefont {Kim}, \citenamefont {Oh}, \citenamefont {Park}, \citenamefont
  {Durairaj}, \citenamefont {Cao},\ and\ \citenamefont
  {Rotenberg}}]{PhysRevLett.101.076402}%
  \BibitemOpen
  \bibfield  {author} {\bibinfo {author} {\bibfnamefont {B.~J.}\ \bibnamefont
  {Kim}}, \bibinfo {author} {\bibfnamefont {H.}~\bibnamefont {Jin}}, \bibinfo
  {author} {\bibfnamefont {S.~J.}\ \bibnamefont {Moon}}, \bibinfo {author}
  {\bibfnamefont {J.-Y.}\ \bibnamefont {Kim}}, \bibinfo {author} {\bibfnamefont
  {B.-G.}\ \bibnamefont {Park}}, \bibinfo {author} {\bibfnamefont {C.~S.}\
  \bibnamefont {Leem}}, \bibinfo {author} {\bibfnamefont {J.}~\bibnamefont
  {Yu}}, \bibinfo {author} {\bibfnamefont {T.~W.}\ \bibnamefont {Noh}},
  \bibinfo {author} {\bibfnamefont {C.}~\bibnamefont {Kim}}, \bibinfo {author}
  {\bibfnamefont {S.-J.}\ \bibnamefont {Oh}}, \bibinfo {author} {\bibfnamefont
  {J.-H.}\ \bibnamefont {Park}}, \bibinfo {author} {\bibfnamefont
  {V.}~\bibnamefont {Durairaj}}, \bibinfo {author} {\bibfnamefont
  {G.}~\bibnamefont {Cao}}, \ and\ \bibinfo {author} {\bibfnamefont
  {E.}~\bibnamefont {Rotenberg}},\ }\href {\doibase
  10.1103/PhysRevLett.101.076402} {\bibfield  {journal} {\bibinfo  {journal}
  {Phys. Rev. Lett.}\ }\textbf {\bibinfo {volume} {101}},\ \bibinfo {pages}
  {076402} (\bibinfo {year} {2008})}\BibitemShut {NoStop}%
\bibitem [{\citenamefont {Jackeli}\ and\ \citenamefont
  {Khaliullin}(2009)}]{PhysRevLett.102.017205}%
  \BibitemOpen
  \bibfield  {author} {\bibinfo {author} {\bibfnamefont {G.}~\bibnamefont
  {Jackeli}}\ and\ \bibinfo {author} {\bibfnamefont {G.}~\bibnamefont
  {Khaliullin}},\ }\href {\doibase 10.1103/PhysRevLett.102.017205} {\bibfield
  {journal} {\bibinfo  {journal} {Phys. Rev. Lett.}\ }\textbf {\bibinfo
  {volume} {102}},\ \bibinfo {pages} {017205} (\bibinfo {year}
  {2009})}\BibitemShut {NoStop}%
\bibitem [{\citenamefont {Liu}\ \emph {et~al.}(2015{\natexlab{a}})\citenamefont
  {Liu}, \citenamefont {Khmelevskyi}, \citenamefont {Kim}, \citenamefont
  {Marsman}, \citenamefont {Li}, \citenamefont {Chen}, \citenamefont {Sarma},
  \citenamefont {Kresse},\ and\ \citenamefont
  {Franchini}}]{PhysRevB.92.054428}%
  \BibitemOpen
  \bibfield  {author} {\bibinfo {author} {\bibfnamefont {P.}~\bibnamefont
  {Liu}}, \bibinfo {author} {\bibfnamefont {S.}~\bibnamefont {Khmelevskyi}},
  \bibinfo {author} {\bibfnamefont {B.}~\bibnamefont {Kim}}, \bibinfo {author}
  {\bibfnamefont {M.}~\bibnamefont {Marsman}}, \bibinfo {author} {\bibfnamefont
  {D.}~\bibnamefont {Li}}, \bibinfo {author} {\bibfnamefont {X.-Q.}\
  \bibnamefont {Chen}}, \bibinfo {author} {\bibfnamefont {D.~D.}\ \bibnamefont
  {Sarma}}, \bibinfo {author} {\bibfnamefont {G.}~\bibnamefont {Kresse}}, \
  and\ \bibinfo {author} {\bibfnamefont {C.}~\bibnamefont {Franchini}},\ }\href
  {\doibase 10.1103/PhysRevB.92.054428} {\bibfield  {journal} {\bibinfo
  {journal} {Phys. Rev. B}\ }\textbf {\bibinfo {volume} {92}},\ \bibinfo
  {pages} {054428} (\bibinfo {year} {2015}{\natexlab{a}})}\BibitemShut
  {NoStop}%
\bibitem [{\citenamefont {Yamaura}(2016)}]{YAMAURA201645}%
  \BibitemOpen
  \bibfield  {author} {\bibinfo {author} {\bibfnamefont {K.}~\bibnamefont
  {Yamaura}},\ }\href {\doibase https://doi.org/10.1016/j.jssc.2015.06.037}
  {\bibfield  {journal} {\bibinfo  {journal} {Journal of Solid State
  Chemistry}\ }\textbf {\bibinfo {volume} {236}},\ \bibinfo {pages} {45 }
  (\bibinfo {year} {2016})}\BibitemShut {NoStop}%
\bibitem [{\citenamefont {Shi}\ \emph {et~al.}(2013)\citenamefont {Shi},
  \citenamefont {Guo}, \citenamefont {Wang}, \citenamefont {Princep},
  \citenamefont {Khalyavin}, \citenamefont {Manuel}, \citenamefont {Michiue},
  \citenamefont {Sato}, \citenamefont {Tsuda}, \citenamefont {Yu},
  \citenamefont {Arai}, \citenamefont {Shirako}, \citenamefont {Akaogi},
  \citenamefont {Wang}, \citenamefont {Yamaura},\ and\ \citenamefont
  {Boothroyd}}]{Shi2013}%
  \BibitemOpen
  \bibfield  {author} {\bibinfo {author} {\bibfnamefont {Y.}~\bibnamefont
  {Shi}}, \bibinfo {author} {\bibfnamefont {Y.}~\bibnamefont {Guo}}, \bibinfo
  {author} {\bibfnamefont {X.}~\bibnamefont {Wang}}, \bibinfo {author}
  {\bibfnamefont {A.~J.}\ \bibnamefont {Princep}}, \bibinfo {author}
  {\bibfnamefont {D.}~\bibnamefont {Khalyavin}}, \bibinfo {author}
  {\bibfnamefont {P.}~\bibnamefont {Manuel}}, \bibinfo {author} {\bibfnamefont
  {Y.}~\bibnamefont {Michiue}}, \bibinfo {author} {\bibfnamefont
  {A.}~\bibnamefont {Sato}}, \bibinfo {author} {\bibfnamefont {K.}~\bibnamefont
  {Tsuda}}, \bibinfo {author} {\bibfnamefont {S.}~\bibnamefont {Yu}}, \bibinfo
  {author} {\bibfnamefont {M.}~\bibnamefont {Arai}}, \bibinfo {author}
  {\bibfnamefont {Y.}~\bibnamefont {Shirako}}, \bibinfo {author} {\bibfnamefont
  {M.}~\bibnamefont {Akaogi}}, \bibinfo {author} {\bibfnamefont
  {N.}~\bibnamefont {Wang}}, \bibinfo {author} {\bibfnamefont {K.}~\bibnamefont
  {Yamaura}}, \ and\ \bibinfo {author} {\bibfnamefont {A.~T.}\ \bibnamefont
  {Boothroyd}},\ }\href {https://www.nature.com/articles/nmat3754} {\bibfield
  {journal} {\bibinfo  {journal} {Nature materials}\ }\textbf {\bibinfo
  {volume} {12}},\ \bibinfo {pages} {1024} (\bibinfo {year}
  {2013})}\BibitemShut {NoStop}%
\bibitem [{\citenamefont {Lu}\ \emph {et~al.}(2019)\citenamefont {Lu},
  \citenamefont {Chen}, \citenamefont {Luo}, \citenamefont {\'I\~niguez},
  \citenamefont {Bellaiche},\ and\ \citenamefont
  {Xiang}}]{PhysRevLett.122.227601}%
  \BibitemOpen
  \bibfield  {author} {\bibinfo {author} {\bibfnamefont {J.}~\bibnamefont
  {Lu}}, \bibinfo {author} {\bibfnamefont {G.}~\bibnamefont {Chen}}, \bibinfo
  {author} {\bibfnamefont {W.}~\bibnamefont {Luo}}, \bibinfo {author}
  {\bibfnamefont {J.}~\bibnamefont {\'I\~niguez}}, \bibinfo {author}
  {\bibfnamefont {L.}~\bibnamefont {Bellaiche}}, \ and\ \bibinfo {author}
  {\bibfnamefont {H.}~\bibnamefont {Xiang}},\ }\href {\doibase
  10.1103/PhysRevLett.122.227601} {\bibfield  {journal} {\bibinfo  {journal}
  {Phys. Rev. Lett.}\ }\textbf {\bibinfo {volume} {122}},\ \bibinfo {pages}
  {227601} (\bibinfo {year} {2019})}\BibitemShut {NoStop}%
\bibitem [{\citenamefont {Shi}\ \emph {et~al.}(2009)\citenamefont {Shi},
  \citenamefont {Guo}, \citenamefont {Yu}, \citenamefont {Arai}, \citenamefont
  {Belik}, \citenamefont {Sato}, \citenamefont {Yamaura}, \citenamefont
  {Takayama-Muromachi}, \citenamefont {Tian}, \citenamefont {Yang},
  \citenamefont {Li}, \citenamefont {Varga}, \citenamefont {Mitchell},\ and\
  \citenamefont {Okamoto}}]{PhysRevB.80.161104}%
  \BibitemOpen
  \bibfield  {author} {\bibinfo {author} {\bibfnamefont {Y.~G.}\ \bibnamefont
  {Shi}}, \bibinfo {author} {\bibfnamefont {Y.~F.}\ \bibnamefont {Guo}},
  \bibinfo {author} {\bibfnamefont {S.}~\bibnamefont {Yu}}, \bibinfo {author}
  {\bibfnamefont {M.}~\bibnamefont {Arai}}, \bibinfo {author} {\bibfnamefont
  {A.~A.}\ \bibnamefont {Belik}}, \bibinfo {author} {\bibfnamefont
  {A.}~\bibnamefont {Sato}}, \bibinfo {author} {\bibfnamefont {K.}~\bibnamefont
  {Yamaura}}, \bibinfo {author} {\bibfnamefont {E.}~\bibnamefont
  {Takayama-Muromachi}}, \bibinfo {author} {\bibfnamefont {H.~F.}\ \bibnamefont
  {Tian}}, \bibinfo {author} {\bibfnamefont {H.~X.}\ \bibnamefont {Yang}},
  \bibinfo {author} {\bibfnamefont {J.~Q.}\ \bibnamefont {Li}}, \bibinfo
  {author} {\bibfnamefont {T.}~\bibnamefont {Varga}}, \bibinfo {author}
  {\bibfnamefont {J.~F.}\ \bibnamefont {Mitchell}}, \ and\ \bibinfo {author}
  {\bibfnamefont {S.}~\bibnamefont {Okamoto}},\ }\href {\doibase
  10.1103/PhysRevB.80.161104} {\bibfield  {journal} {\bibinfo  {journal} {Phys.
  Rev. B}\ }\textbf {\bibinfo {volume} {80}},\ \bibinfo {pages} {161104}
  (\bibinfo {year} {2009})}\BibitemShut {NoStop}%
\bibitem [{\citenamefont {Kim}\ \emph {et~al.}(2016)\citenamefont {Kim},
  \citenamefont {Liu}, \citenamefont {Erg\"onenc}, \citenamefont {Toschi},
  \citenamefont {Khmelevskyi},\ and\ \citenamefont
  {Franchini}}]{PhysRevB.94.241113}%
  \BibitemOpen
  \bibfield  {author} {\bibinfo {author} {\bibfnamefont {B.}~\bibnamefont
  {Kim}}, \bibinfo {author} {\bibfnamefont {P.}~\bibnamefont {Liu}}, \bibinfo
  {author} {\bibfnamefont {Z.}~\bibnamefont {Erg\"onenc}}, \bibinfo {author}
  {\bibfnamefont {A.}~\bibnamefont {Toschi}}, \bibinfo {author} {\bibfnamefont
  {S.}~\bibnamefont {Khmelevskyi}}, \ and\ \bibinfo {author} {\bibfnamefont
  {C.}~\bibnamefont {Franchini}},\ }\href {\doibase 10.1103/PhysRevB.94.241113}
  {\bibfield  {journal} {\bibinfo  {journal} {Phys. Rev. B}\ }\textbf {\bibinfo
  {volume} {94}},\ \bibinfo {pages} {241113} (\bibinfo {year}
  {2016})}\BibitemShut {NoStop}%
\bibitem [{\citenamefont {Calder}\ \emph {et~al.}(2015)\citenamefont {Calder},
  \citenamefont {Lee}, \citenamefont {Stone}, \citenamefont {Lumsden},
  \citenamefont {Lang}, \citenamefont {Feygenson}, \citenamefont {Zhao},
  \citenamefont {Yan}, \citenamefont {Shi}, \citenamefont {Sun}, \citenamefont
  {Tsujimoto}, \citenamefont {Yamaura},\ and\ \citenamefont
  {Christianson}}]{Calder2015}%
  \BibitemOpen
  \bibfield  {author} {\bibinfo {author} {\bibfnamefont {S.}~\bibnamefont
  {Calder}}, \bibinfo {author} {\bibfnamefont {J.~H.}\ \bibnamefont {Lee}},
  \bibinfo {author} {\bibfnamefont {M.~B.}\ \bibnamefont {Stone}}, \bibinfo
  {author} {\bibfnamefont {M.~D.}\ \bibnamefont {Lumsden}}, \bibinfo {author}
  {\bibfnamefont {J.~C.}\ \bibnamefont {Lang}}, \bibinfo {author}
  {\bibfnamefont {M.}~\bibnamefont {Feygenson}}, \bibinfo {author}
  {\bibfnamefont {Z.}~\bibnamefont {Zhao}}, \bibinfo {author} {\bibfnamefont
  {J.~Q.}\ \bibnamefont {Yan}}, \bibinfo {author} {\bibfnamefont {Y.~G.}\
  \bibnamefont {Shi}}, \bibinfo {author} {\bibfnamefont {Y.~S.}\ \bibnamefont
  {Sun}}, \bibinfo {author} {\bibfnamefont {Y.}~\bibnamefont {Tsujimoto}},
  \bibinfo {author} {\bibfnamefont {K.}~\bibnamefont {Yamaura}}, \ and\
  \bibinfo {author} {\bibfnamefont {A.~D.}\ \bibnamefont {Christianson}},\
  }\href {https://doi.org/10.1038/ncomms9916} {\bibfield  {journal} {\bibinfo
  {journal} {Nat Commun}\ }\textbf {\bibinfo {volume} {6}},\ \bibinfo {pages}
  {8916} (\bibinfo {year} {2015})}\BibitemShut {NoStop}%
\bibitem [{\citenamefont {Xiang}(2014)}]{PhysRevB.90.094108}%
  \BibitemOpen
  \bibfield  {author} {\bibinfo {author} {\bibfnamefont {H.~J.}\ \bibnamefont
  {Xiang}},\ }\href {\doibase 10.1103/PhysRevB.90.094108} {\bibfield  {journal}
  {\bibinfo  {journal} {Phys. Rev. B}\ }\textbf {\bibinfo {volume} {90}},\
  \bibinfo {pages} {094108} (\bibinfo {year} {2014})}\BibitemShut {NoStop}%
\bibitem [{\citenamefont {Sim}\ and\ \citenamefont
  {Kim}(2014)}]{PhysRevB.89.201107}%
  \BibitemOpen
  \bibfield  {author} {\bibinfo {author} {\bibfnamefont {H.}~\bibnamefont
  {Sim}}\ and\ \bibinfo {author} {\bibfnamefont {B.~G.}\ \bibnamefont {Kim}},\
  }\href {\doibase 10.1103/PhysRevB.89.201107} {\bibfield  {journal} {\bibinfo
  {journal} {Phys. Rev. B}\ }\textbf {\bibinfo {volume} {89}},\ \bibinfo
  {pages} {201107} (\bibinfo {year} {2014})}\BibitemShut {NoStop}%
\bibitem [{\citenamefont {Giovannetti}\ and\ \citenamefont
  {Capone}(2014)}]{PhysRevB.90.195113}%
  \BibitemOpen
  \bibfield  {author} {\bibinfo {author} {\bibfnamefont {G.}~\bibnamefont
  {Giovannetti}}\ and\ \bibinfo {author} {\bibfnamefont {M.}~\bibnamefont
  {Capone}},\ }\href {\doibase 10.1103/PhysRevB.90.195113} {\bibfield
  {journal} {\bibinfo  {journal} {Phys. Rev. B}\ }\textbf {\bibinfo {volume}
  {90}},\ \bibinfo {pages} {195113} (\bibinfo {year} {2014})}\BibitemShut
  {NoStop}%
\bibitem [{\citenamefont {Liu}\ \emph {et~al.}(2015{\natexlab{b}})\citenamefont
  {Liu}, \citenamefont {Du}, \citenamefont {Xie}, \citenamefont {Liu},
  \citenamefont {Duan},\ and\ \citenamefont {Wan}}]{PhysRevB.91.064104}%
  \BibitemOpen
  \bibfield  {author} {\bibinfo {author} {\bibfnamefont {H.~M.}\ \bibnamefont
  {Liu}}, \bibinfo {author} {\bibfnamefont {Y.~P.}\ \bibnamefont {Du}},
  \bibinfo {author} {\bibfnamefont {Y.~L.}\ \bibnamefont {Xie}}, \bibinfo
  {author} {\bibfnamefont {J.-M.}\ \bibnamefont {Liu}}, \bibinfo {author}
  {\bibfnamefont {C.-G.}\ \bibnamefont {Duan}}, \ and\ \bibinfo {author}
  {\bibfnamefont {X.}~\bibnamefont {Wan}},\ }\href {\doibase
  10.1103/PhysRevB.91.064104} {\bibfield  {journal} {\bibinfo  {journal} {Phys.
  Rev. B}\ }\textbf {\bibinfo {volume} {91}},\ \bibinfo {pages} {064104}
  (\bibinfo {year} {2015}{\natexlab{b}})}\BibitemShut {NoStop}%
\bibitem [{\citenamefont {Kirschner}\ \emph {et~al.}()\citenamefont
  {Kirschner}, \citenamefont {Lang}, \citenamefont {Pratt}, \citenamefont
  {Lancaster}, \citenamefont {Shi}, \citenamefont {Guo}, \citenamefont
  {Boothroyd},\ and\ \citenamefont {Blundell}}]{doi:10.7566/JPSCP.21.011013}%
  \BibitemOpen
  \bibfield  {author} {\bibinfo {author} {\bibfnamefont {F.~K.~K.}\
  \bibnamefont {Kirschner}}, \bibinfo {author} {\bibfnamefont {F.}~\bibnamefont
  {Lang}}, \bibinfo {author} {\bibfnamefont {F.~L.}\ \bibnamefont {Pratt}},
  \bibinfo {author} {\bibfnamefont {T.}~\bibnamefont {Lancaster}}, \bibinfo
  {author} {\bibfnamefont {Y.}~\bibnamefont {Shi}}, \bibinfo {author}
  {\bibfnamefont {Y.}~\bibnamefont {Guo}}, \bibinfo {author} {\bibfnamefont
  {A.~T.}\ \bibnamefont {Boothroyd}}, \ and\ \bibinfo {author} {\bibfnamefont
  {S.~J.}\ \bibnamefont {Blundell}},\ }\enquote {\bibinfo {title} {Static and
  fluctuating magnetic moments in the ferroelectric metal lioso$_3$},}\ in\
  \href {\doibase 10.7566/JPSCP.21.011013} {\emph {\bibinfo {booktitle}
  {Proceedings of the 14th International Conference on Muon Spin Rotation,
  Relaxation and Resonance ($\mu$SR2017)}}}\BibitemShut {NoStop}%
\bibitem [{\citenamefont {Lo~Vecchio}\ \emph {et~al.}(2013)\citenamefont
  {Lo~Vecchio}, \citenamefont {Perucchi}, \citenamefont {Di~Pietro},
  \citenamefont {Limaj}, \citenamefont {Schade}, \citenamefont {Sun},
  \citenamefont {Arai}, \citenamefont {Yamaura},\ and\ \citenamefont
  {Lupi}}]{LoVecchio2013}%
  \BibitemOpen
  \bibfield  {author} {\bibinfo {author} {\bibfnamefont {I.}~\bibnamefont
  {Lo~Vecchio}}, \bibinfo {author} {\bibfnamefont {A.}~\bibnamefont
  {Perucchi}}, \bibinfo {author} {\bibfnamefont {P.}~\bibnamefont {Di~Pietro}},
  \bibinfo {author} {\bibfnamefont {O.}~\bibnamefont {Limaj}}, \bibinfo
  {author} {\bibfnamefont {U.}~\bibnamefont {Schade}}, \bibinfo {author}
  {\bibfnamefont {Y.}~\bibnamefont {Sun}}, \bibinfo {author} {\bibfnamefont
  {M.}~\bibnamefont {Arai}}, \bibinfo {author} {\bibfnamefont {K.}~\bibnamefont
  {Yamaura}}, \ and\ \bibinfo {author} {\bibfnamefont {S.}~\bibnamefont
  {Lupi}},\ }\href {\doibase 10.1038/srep02990} {\bibfield  {journal} {\bibinfo
   {journal} {Scientific Reports}\ }\textbf {\bibinfo {volume} {3}},\ \bibinfo
  {pages} {2990} (\bibinfo {year} {2013})}\BibitemShut {NoStop}%
\bibitem [{\citenamefont {Calder}\ \emph {et~al.}(2012)\citenamefont {Calder},
  \citenamefont {Garlea}, \citenamefont {McMorrow}, \citenamefont {Lumsden},
  \citenamefont {Stone}, \citenamefont {Lang}, \citenamefont {Kim},
  \citenamefont {Schlueter}, \citenamefont {Shi}, \citenamefont {Yamaura},
  \citenamefont {Sun}, \citenamefont {Tsujimoto},\ and\ \citenamefont
  {Christianson}}]{PhysRevLett.108.257209}%
  \BibitemOpen
  \bibfield  {author} {\bibinfo {author} {\bibfnamefont {S.}~\bibnamefont
  {Calder}}, \bibinfo {author} {\bibfnamefont {V.~O.}\ \bibnamefont {Garlea}},
  \bibinfo {author} {\bibfnamefont {D.~F.}\ \bibnamefont {McMorrow}}, \bibinfo
  {author} {\bibfnamefont {M.~D.}\ \bibnamefont {Lumsden}}, \bibinfo {author}
  {\bibfnamefont {M.~B.}\ \bibnamefont {Stone}}, \bibinfo {author}
  {\bibfnamefont {J.~C.}\ \bibnamefont {Lang}}, \bibinfo {author}
  {\bibfnamefont {J.-W.}\ \bibnamefont {Kim}}, \bibinfo {author} {\bibfnamefont
  {J.~A.}\ \bibnamefont {Schlueter}}, \bibinfo {author} {\bibfnamefont {Y.~G.}\
  \bibnamefont {Shi}}, \bibinfo {author} {\bibfnamefont {K.}~\bibnamefont
  {Yamaura}}, \bibinfo {author} {\bibfnamefont {Y.~S.}\ \bibnamefont {Sun}},
  \bibinfo {author} {\bibfnamefont {Y.}~\bibnamefont {Tsujimoto}}, \ and\
  \bibinfo {author} {\bibfnamefont {A.~D.}\ \bibnamefont {Christianson}},\
  }\href {\doibase 10.1103/PhysRevLett.108.257209} {\bibfield  {journal}
  {\bibinfo  {journal} {Phys. Rev. Lett.}\ }\textbf {\bibinfo {volume} {108}},\
  \bibinfo {pages} {257209} (\bibinfo {year} {2012})}\BibitemShut {NoStop}%
\bibitem [{\citenamefont {Du}\ \emph {et~al.}(2012)\citenamefont {Du},
  \citenamefont {Wan}, \citenamefont {Sheng}, \citenamefont {Dong},\ and\
  \citenamefont {Savrasov}}]{PhysRevB.85.174424}%
  \BibitemOpen
  \bibfield  {author} {\bibinfo {author} {\bibfnamefont {Y.}~\bibnamefont
  {Du}}, \bibinfo {author} {\bibfnamefont {X.}~\bibnamefont {Wan}}, \bibinfo
  {author} {\bibfnamefont {L.}~\bibnamefont {Sheng}}, \bibinfo {author}
  {\bibfnamefont {J.}~\bibnamefont {Dong}}, \ and\ \bibinfo {author}
  {\bibfnamefont {S.~Y.}\ \bibnamefont {Savrasov}},\ }\href {\doibase
  10.1103/PhysRevB.85.174424} {\bibfield  {journal} {\bibinfo  {journal} {Phys.
  Rev. B}\ }\textbf {\bibinfo {volume} {85}},\ \bibinfo {pages} {174424}
  (\bibinfo {year} {2012})}\BibitemShut {NoStop}%
\bibitem [{\citenamefont {Jung}\ \emph {et~al.}(2013)\citenamefont {Jung},
  \citenamefont {Song}, \citenamefont {Lee},\ and\ \citenamefont
  {Pickett}}]{PhysRevB.87.115119}%
  \BibitemOpen
  \bibfield  {author} {\bibinfo {author} {\bibfnamefont {M.-C.}\ \bibnamefont
  {Jung}}, \bibinfo {author} {\bibfnamefont {Y.-J.}\ \bibnamefont {Song}},
  \bibinfo {author} {\bibfnamefont {K.-W.}\ \bibnamefont {Lee}}, \ and\
  \bibinfo {author} {\bibfnamefont {W.~E.}\ \bibnamefont {Pickett}},\ }\href
  {\doibase 10.1103/PhysRevB.87.115119} {\bibfield  {journal} {\bibinfo
  {journal} {Phys. Rev. B}\ }\textbf {\bibinfo {volume} {87}},\ \bibinfo
  {pages} {115119} (\bibinfo {year} {2013})}\BibitemShut {NoStop}%
\bibitem [{\citenamefont {Vale}\ \emph
  {et~al.}(2018{\natexlab{a}})\citenamefont {Vale}, \citenamefont {Calder},
  \citenamefont {Donnerer}, \citenamefont {Pincini}, \citenamefont {Shi},
  \citenamefont {Tsujimoto}, \citenamefont {Yamaura}, \citenamefont {Sala},
  \citenamefont {van~den Brink}, \citenamefont {Christianson},\ and\
  \citenamefont {McMorrow}}]{PhysRevLett.120.227203}%
  \BibitemOpen
  \bibfield  {author} {\bibinfo {author} {\bibfnamefont {J.~G.}\ \bibnamefont
  {Vale}}, \bibinfo {author} {\bibfnamefont {S.}~\bibnamefont {Calder}},
  \bibinfo {author} {\bibfnamefont {C.}~\bibnamefont {Donnerer}}, \bibinfo
  {author} {\bibfnamefont {D.}~\bibnamefont {Pincini}}, \bibinfo {author}
  {\bibfnamefont {Y.~G.}\ \bibnamefont {Shi}}, \bibinfo {author} {\bibfnamefont
  {Y.}~\bibnamefont {Tsujimoto}}, \bibinfo {author} {\bibfnamefont
  {K.}~\bibnamefont {Yamaura}}, \bibinfo {author} {\bibfnamefont {M.~M.}\
  \bibnamefont {Sala}}, \bibinfo {author} {\bibfnamefont {J.}~\bibnamefont
  {van~den Brink}}, \bibinfo {author} {\bibfnamefont {A.~D.}\ \bibnamefont
  {Christianson}}, \ and\ \bibinfo {author} {\bibfnamefont {D.~F.}\
  \bibnamefont {McMorrow}},\ }\href {\doibase 10.1103/PhysRevLett.120.227203}
  {\bibfield  {journal} {\bibinfo  {journal} {Phys. Rev. Lett.}\ }\textbf
  {\bibinfo {volume} {120}},\ \bibinfo {pages} {227203} (\bibinfo {year}
  {2018}{\natexlab{a}})}\BibitemShut {NoStop}%
\bibitem [{\citenamefont {Vale}\ \emph
  {et~al.}(2018{\natexlab{b}})\citenamefont {Vale}, \citenamefont {Calder},
  \citenamefont {Donnerer}, \citenamefont {Pincini}, \citenamefont {Shi},
  \citenamefont {Tsujimoto}, \citenamefont {Yamaura}, \citenamefont
  {Moretti~Sala}, \citenamefont {van~den Brink}, \citenamefont {Christianson},\
  and\ \citenamefont {McMorrow}}]{PhysRevB.97.184429}%
  \BibitemOpen
  \bibfield  {author} {\bibinfo {author} {\bibfnamefont {J.~G.}\ \bibnamefont
  {Vale}}, \bibinfo {author} {\bibfnamefont {S.}~\bibnamefont {Calder}},
  \bibinfo {author} {\bibfnamefont {C.}~\bibnamefont {Donnerer}}, \bibinfo
  {author} {\bibfnamefont {D.}~\bibnamefont {Pincini}}, \bibinfo {author}
  {\bibfnamefont {Y.~G.}\ \bibnamefont {Shi}}, \bibinfo {author} {\bibfnamefont
  {Y.}~\bibnamefont {Tsujimoto}}, \bibinfo {author} {\bibfnamefont
  {K.}~\bibnamefont {Yamaura}}, \bibinfo {author} {\bibfnamefont
  {M.}~\bibnamefont {Moretti~Sala}}, \bibinfo {author} {\bibfnamefont
  {J.}~\bibnamefont {van~den Brink}}, \bibinfo {author} {\bibfnamefont {A.~D.}\
  \bibnamefont {Christianson}}, \ and\ \bibinfo {author} {\bibfnamefont
  {D.~F.}\ \bibnamefont {McMorrow}},\ }\href {\doibase
  10.1103/PhysRevB.97.184429} {\bibfield  {journal} {\bibinfo  {journal} {Phys.
  Rev. B}\ }\textbf {\bibinfo {volume} {97}},\ \bibinfo {pages} {184429}
  (\bibinfo {year} {2018}{\natexlab{b}})}\BibitemShut {NoStop}%
\bibitem [{\citenamefont {Aryasetiawan}\ \emph {et~al.}(2006)\citenamefont
  {Aryasetiawan}, \citenamefont {Karlsson}, \citenamefont {Jepsen},\ and\
  \citenamefont {Sch\"onberger}}]{PhysRevB.74.125106}%
  \BibitemOpen
  \bibfield  {author} {\bibinfo {author} {\bibfnamefont {F.}~\bibnamefont
  {Aryasetiawan}}, \bibinfo {author} {\bibfnamefont {K.}~\bibnamefont
  {Karlsson}}, \bibinfo {author} {\bibfnamefont {O.}~\bibnamefont {Jepsen}}, \
  and\ \bibinfo {author} {\bibfnamefont {U.}~\bibnamefont {Sch\"onberger}},\
  }\href {\doibase 10.1103/PhysRevB.74.125106} {\bibfield  {journal} {\bibinfo
  {journal} {Phys. Rev. B}\ }\textbf {\bibinfo {volume} {74}},\ \bibinfo
  {pages} {125106} (\bibinfo {year} {2006})}\BibitemShut {NoStop}%
\bibitem [{\citenamefont {Springer}\ \emph {et~al.}(2019)\citenamefont
  {Springer}, \citenamefont {Kim}, \citenamefont {Liu}, \citenamefont
  {Khmelevskyi}, \citenamefont {Capone}, \citenamefont {Sangiovanni},
  \citenamefont {Franchini},\ and\ \citenamefont {Toschi}}]{Daniel2019}%
  \BibitemOpen
  \bibfield  {author} {\bibinfo {author} {\bibfnamefont {D.}~\bibnamefont
  {Springer}}, \bibinfo {author} {\bibfnamefont {B.}~\bibnamefont {Kim}},
  \bibinfo {author} {\bibfnamefont {P.}~\bibnamefont {Liu}}, \bibinfo {author}
  {\bibfnamefont {S.}~\bibnamefont {Khmelevskyi}}, \bibinfo {author}
  {\bibfnamefont {M.}~\bibnamefont {Capone}}, \bibinfo {author} {\bibfnamefont
  {G.}~\bibnamefont {Sangiovanni}}, \bibinfo {author} {\bibfnamefont
  {C.}~\bibnamefont {Franchini}}, \ and\ \bibinfo {author} {\bibfnamefont
  {A.}~\bibnamefont {Toschi}},\ }\href {https://arxiv.org/abs/1910.05151}
  {\bibfield  {journal} {\bibinfo  {journal} {arXiv:1910.05151v1}\ } (\bibinfo
  {year} {2019})}\BibitemShut {NoStop}%
\bibitem [{\citenamefont {Lo~Vecchio}\ \emph {et~al.}(2016)\citenamefont
  {Lo~Vecchio}, \citenamefont {Giovannetti}, \citenamefont {Autore},
  \citenamefont {Di~Pietro}, \citenamefont {Perucchi}, \citenamefont {He},
  \citenamefont {Yamaura}, \citenamefont {Capone},\ and\ \citenamefont
  {Lupi}}]{PhysRevB.93.161113}%
  \BibitemOpen
  \bibfield  {author} {\bibinfo {author} {\bibfnamefont {I.}~\bibnamefont
  {Lo~Vecchio}}, \bibinfo {author} {\bibfnamefont {G.}~\bibnamefont
  {Giovannetti}}, \bibinfo {author} {\bibfnamefont {M.}~\bibnamefont {Autore}},
  \bibinfo {author} {\bibfnamefont {P.}~\bibnamefont {Di~Pietro}}, \bibinfo
  {author} {\bibfnamefont {A.}~\bibnamefont {Perucchi}}, \bibinfo {author}
  {\bibfnamefont {J.}~\bibnamefont {He}}, \bibinfo {author} {\bibfnamefont
  {K.}~\bibnamefont {Yamaura}}, \bibinfo {author} {\bibfnamefont
  {M.}~\bibnamefont {Capone}}, \ and\ \bibinfo {author} {\bibfnamefont
  {S.}~\bibnamefont {Lupi}},\ }\href {\doibase 10.1103/PhysRevB.93.161113}
  {\bibfield  {journal} {\bibinfo  {journal} {Phys. Rev. B}\ }\textbf {\bibinfo
  {volume} {93}},\ \bibinfo {pages} {161113} (\bibinfo {year}
  {2016})}\BibitemShut {NoStop}%
\bibitem [{\citenamefont {Baldassarre}\ \emph {et~al.}(2008)\citenamefont
  {Baldassarre}, \citenamefont {Perucchi}, \citenamefont {Nicoletti},
  \citenamefont {Toschi}, \citenamefont {Sangiovanni}, \citenamefont {Held},
  \citenamefont {Capone}, \citenamefont {Ortolani}, \citenamefont {Malavasi},
  \citenamefont {Marsi}, \citenamefont {Metcalf}, \citenamefont {Postorino},\
  and\ \citenamefont {Lupi}}]{PhysRevB.77.113107}%
  \BibitemOpen
  \bibfield  {author} {\bibinfo {author} {\bibfnamefont {L.}~\bibnamefont
  {Baldassarre}}, \bibinfo {author} {\bibfnamefont {A.}~\bibnamefont
  {Perucchi}}, \bibinfo {author} {\bibfnamefont {D.}~\bibnamefont {Nicoletti}},
  \bibinfo {author} {\bibfnamefont {A.}~\bibnamefont {Toschi}}, \bibinfo
  {author} {\bibfnamefont {G.}~\bibnamefont {Sangiovanni}}, \bibinfo {author}
  {\bibfnamefont {K.}~\bibnamefont {Held}}, \bibinfo {author} {\bibfnamefont
  {M.}~\bibnamefont {Capone}}, \bibinfo {author} {\bibfnamefont
  {M.}~\bibnamefont {Ortolani}}, \bibinfo {author} {\bibfnamefont
  {L.}~\bibnamefont {Malavasi}}, \bibinfo {author} {\bibfnamefont
  {M.}~\bibnamefont {Marsi}}, \bibinfo {author} {\bibfnamefont
  {P.}~\bibnamefont {Metcalf}}, \bibinfo {author} {\bibfnamefont
  {P.}~\bibnamefont {Postorino}}, \ and\ \bibinfo {author} {\bibfnamefont
  {S.}~\bibnamefont {Lupi}},\ }\href {\doibase 10.1103/PhysRevB.77.113107}
  {\bibfield  {journal} {\bibinfo  {journal} {Phys. Rev. B}\ }\textbf {\bibinfo
  {volume} {77}},\ \bibinfo {pages} {113107} (\bibinfo {year}
  {2008})}\BibitemShut {NoStop}%
\bibitem [{\citenamefont {Trimarchi}\ \emph {et~al.}(2018)\citenamefont
  {Trimarchi}, \citenamefont {Wang},\ and\ \citenamefont
  {Zunger}}]{PhysRevB.97.035107}%
  \BibitemOpen
  \bibfield  {author} {\bibinfo {author} {\bibfnamefont {G.}~\bibnamefont
  {Trimarchi}}, \bibinfo {author} {\bibfnamefont {Z.}~\bibnamefont {Wang}}, \
  and\ \bibinfo {author} {\bibfnamefont {A.}~\bibnamefont {Zunger}},\ }\href
  {\doibase 10.1103/PhysRevB.97.035107} {\bibfield  {journal} {\bibinfo
  {journal} {Phys. Rev. B}\ }\textbf {\bibinfo {volume} {97}},\ \bibinfo
  {pages} {035107} (\bibinfo {year} {2018})}\BibitemShut {NoStop}%
\bibitem [{\citenamefont {Bl\"ochl}(1994)}]{PhysRevB.50.17953}%
  \BibitemOpen
  \bibfield  {author} {\bibinfo {author} {\bibfnamefont {P.~E.}\ \bibnamefont
  {Bl\"ochl}},\ }\href {\doibase 10.1103/PhysRevB.50.17953} {\bibfield
  {journal} {\bibinfo  {journal} {Phys. Rev. B}\ }\textbf {\bibinfo {volume}
  {50}},\ \bibinfo {pages} {17953} (\bibinfo {year} {1994})}\BibitemShut
  {NoStop}%
\bibitem [{\citenamefont {Kresse}\ and\ \citenamefont
  {Hafner}(1993)}]{PhysRevB.47.558}%
  \BibitemOpen
  \bibfield  {author} {\bibinfo {author} {\bibfnamefont {G.}~\bibnamefont
  {Kresse}}\ and\ \bibinfo {author} {\bibfnamefont {J.}~\bibnamefont
  {Hafner}},\ }\href {\doibase 10.1103/PhysRevB.47.558} {\bibfield  {journal}
  {\bibinfo  {journal} {Phys. Rev. B}\ }\textbf {\bibinfo {volume} {47}},\
  \bibinfo {pages} {558} (\bibinfo {year} {1993})}\BibitemShut {NoStop}%
\bibitem [{\citenamefont {Kresse}\ and\ \citenamefont
  {Furthm\"uller}(1996)}]{PhysRevB.54.11169}%
  \BibitemOpen
  \bibfield  {author} {\bibinfo {author} {\bibfnamefont {G.}~\bibnamefont
  {Kresse}}\ and\ \bibinfo {author} {\bibfnamefont {J.}~\bibnamefont
  {Furthm\"uller}},\ }\href {\doibase 10.1103/PhysRevB.54.11169} {\bibfield
  {journal} {\bibinfo  {journal} {Phys. Rev. B}\ }\textbf {\bibinfo {volume}
  {54}},\ \bibinfo {pages} {11169} (\bibinfo {year} {1996})}\BibitemShut
  {NoStop}%
\bibitem [{\citenamefont {ISOTROPY}()}]{ISOTROPY}%
  \BibitemOpen
  \bibfield  {author} {\bibinfo {author} {\bibnamefont {ISOTROPY}},\ }\href
  {http://stokes.byu.edu/isotropy.html} {\emph {\bibinfo {title}
  {http://stokes.byu.edu/isotropy.html}}}\BibitemShut {NoStop}%
\bibitem [{\citenamefont {Orobengoa}\ \emph {et~al.}(2009)\citenamefont
  {Orobengoa}, \citenamefont {Capillas}, \citenamefont {Aroyo},\ and\
  \citenamefont {Perez-Mato}}]{Orobengoa:ks5225}%
  \BibitemOpen
  \bibfield  {author} {\bibinfo {author} {\bibfnamefont {D.}~\bibnamefont
  {Orobengoa}}, \bibinfo {author} {\bibfnamefont {C.}~\bibnamefont {Capillas}},
  \bibinfo {author} {\bibfnamefont {M.~I.}\ \bibnamefont {Aroyo}}, \ and\
  \bibinfo {author} {\bibfnamefont {J.~M.}\ \bibnamefont {Perez-Mato}},\ }\href
  {\doibase 10.1107/S0021889809028064} {\bibfield  {journal} {\bibinfo
  {journal} {Journal of Applied Crystallography}\ }\textbf {\bibinfo {volume}
  {42}},\ \bibinfo {pages} {820} (\bibinfo {year} {2009})}\BibitemShut
  {NoStop}%
\bibitem [{\citenamefont {Perdew}\ and\ \citenamefont
  {Zunger}(1981)}]{PhysRevB.23.5048}%
  \BibitemOpen
  \bibfield  {author} {\bibinfo {author} {\bibfnamefont {J.~P.}\ \bibnamefont
  {Perdew}}\ and\ \bibinfo {author} {\bibfnamefont {A.}~\bibnamefont
  {Zunger}},\ }\href {\doibase 10.1103/PhysRevB.23.5048} {\bibfield  {journal}
  {\bibinfo  {journal} {Phys. Rev. B}\ }\textbf {\bibinfo {volume} {23}},\
  \bibinfo {pages} {5048} (\bibinfo {year} {1981})}\BibitemShut {NoStop}%
\bibitem [{\citenamefont {Ceperley}\ and\ \citenamefont
  {Alder}(1980)}]{PhysRevLett.45.566}%
  \BibitemOpen
  \bibfield  {author} {\bibinfo {author} {\bibfnamefont {D.~M.}\ \bibnamefont
  {Ceperley}}\ and\ \bibinfo {author} {\bibfnamefont {B.~J.}\ \bibnamefont
  {Alder}},\ }\href {\doibase 10.1103/PhysRevLett.45.566} {\bibfield  {journal}
  {\bibinfo  {journal} {Phys. Rev. Lett.}\ }\textbf {\bibinfo {volume} {45}},\
  \bibinfo {pages} {566} (\bibinfo {year} {1980})}\BibitemShut {NoStop}%
\bibitem [{\citenamefont {Perdew}\ \emph {et~al.}(1996)\citenamefont {Perdew},
  \citenamefont {Burke},\ and\ \citenamefont
  {Ernzerhof}}]{PhysRevLett.77.3865}%
  \BibitemOpen
  \bibfield  {author} {\bibinfo {author} {\bibfnamefont {J.~P.}\ \bibnamefont
  {Perdew}}, \bibinfo {author} {\bibfnamefont {K.}~\bibnamefont {Burke}}, \
  and\ \bibinfo {author} {\bibfnamefont {M.}~\bibnamefont {Ernzerhof}},\ }\href
  {\doibase 10.1103/PhysRevLett.77.3865} {\bibfield  {journal} {\bibinfo
  {journal} {Phys. Rev. Lett.}\ }\textbf {\bibinfo {volume} {77}},\ \bibinfo
  {pages} {3865} (\bibinfo {year} {1996})}\BibitemShut {NoStop}%
\bibitem [{\citenamefont {Perdew}\ \emph {et~al.}(2008)\citenamefont {Perdew},
  \citenamefont {Ruzsinszky}, \citenamefont {Csonka}, \citenamefont {Vydrov},
  \citenamefont {Scuseria}, \citenamefont {Constantin}, \citenamefont {Zhou},\
  and\ \citenamefont {Burke}}]{PhysRevLett.100.136406}%
  \BibitemOpen
  \bibfield  {author} {\bibinfo {author} {\bibfnamefont {J.~P.}\ \bibnamefont
  {Perdew}}, \bibinfo {author} {\bibfnamefont {A.}~\bibnamefont {Ruzsinszky}},
  \bibinfo {author} {\bibfnamefont {G.~I.}\ \bibnamefont {Csonka}}, \bibinfo
  {author} {\bibfnamefont {O.~A.}\ \bibnamefont {Vydrov}}, \bibinfo {author}
  {\bibfnamefont {G.~E.}\ \bibnamefont {Scuseria}}, \bibinfo {author}
  {\bibfnamefont {L.~A.}\ \bibnamefont {Constantin}}, \bibinfo {author}
  {\bibfnamefont {X.}~\bibnamefont {Zhou}}, \ and\ \bibinfo {author}
  {\bibfnamefont {K.}~\bibnamefont {Burke}},\ }\href {\doibase
  10.1103/PhysRevLett.100.136406} {\bibfield  {journal} {\bibinfo  {journal}
  {Phys. Rev. Lett.}\ }\textbf {\bibinfo {volume} {100}},\ \bibinfo {pages}
  {136406} (\bibinfo {year} {2008})}\BibitemShut {NoStop}%
\bibitem [{\citenamefont {Sun}\ \emph {et~al.}(2015)\citenamefont {Sun},
  \citenamefont {Ruzsinszky},\ and\ \citenamefont
  {Perdew}}]{PhysRevLett.115.036402}%
  \BibitemOpen
  \bibfield  {author} {\bibinfo {author} {\bibfnamefont {J.}~\bibnamefont
  {Sun}}, \bibinfo {author} {\bibfnamefont {A.}~\bibnamefont {Ruzsinszky}}, \
  and\ \bibinfo {author} {\bibfnamefont {J.~P.}\ \bibnamefont {Perdew}},\
  }\href {\doibase 10.1103/PhysRevLett.115.036402} {\bibfield  {journal}
  {\bibinfo  {journal} {Phys. Rev. Lett.}\ }\textbf {\bibinfo {volume} {115}},\
  \bibinfo {pages} {036402} (\bibinfo {year} {2015})}\BibitemShut {NoStop}%
\bibitem [{\citenamefont {Krukau}\ \emph {et~al.}(2006)\citenamefont {Krukau},
  \citenamefont {Vydrov}, \citenamefont {Izmaylov},\ and\ \citenamefont
  {Scuseria}}]{doi:10.1063/1.2404663}%
  \BibitemOpen
  \bibfield  {author} {\bibinfo {author} {\bibfnamefont {A.~V.}\ \bibnamefont
  {Krukau}}, \bibinfo {author} {\bibfnamefont {O.~A.}\ \bibnamefont {Vydrov}},
  \bibinfo {author} {\bibfnamefont {A.~F.}\ \bibnamefont {Izmaylov}}, \ and\
  \bibinfo {author} {\bibfnamefont {G.~E.}\ \bibnamefont {Scuseria}},\ }\href
  {\doibase 10.1063/1.2404663} {\bibfield  {journal} {\bibinfo  {journal} {The
  Journal of Chemical Physics}\ }\textbf {\bibinfo {volume} {125}},\ \bibinfo
  {pages} {224106} (\bibinfo {year} {2006})}\BibitemShut {NoStop}%
\bibitem [{\citenamefont {Dudarev}\ \emph {et~al.}(2019)\citenamefont
  {Dudarev}, \citenamefont {Liu}, \citenamefont {Andersson}, \citenamefont
  {Stanek}, \citenamefont {Ozaki},\ and\ \citenamefont
  {Franchini}}]{PhysRevMaterials.3.083802}%
  \BibitemOpen
  \bibfield  {author} {\bibinfo {author} {\bibfnamefont {S.~L.}\ \bibnamefont
  {Dudarev}}, \bibinfo {author} {\bibfnamefont {P.}~\bibnamefont {Liu}},
  \bibinfo {author} {\bibfnamefont {D.~A.}\ \bibnamefont {Andersson}}, \bibinfo
  {author} {\bibfnamefont {C.~R.}\ \bibnamefont {Stanek}}, \bibinfo {author}
  {\bibfnamefont {T.}~\bibnamefont {Ozaki}}, \ and\ \bibinfo {author}
  {\bibfnamefont {C.}~\bibnamefont {Franchini}},\ }\href {\doibase
  10.1103/PhysRevMaterials.3.083802} {\bibfield  {journal} {\bibinfo  {journal}
  {Phys. Rev. Materials}\ }\textbf {\bibinfo {volume} {3}},\ \bibinfo {pages}
  {083802} (\bibinfo {year} {2019})}\BibitemShut {NoStop}%
\bibitem [{\citenamefont {Payne}\ \emph {et~al.}(1992)\citenamefont {Payne},
  \citenamefont {Teter}, \citenamefont {Allan}, \citenamefont {Arias},\ and\
  \citenamefont {Joannopoulos}}]{RevModPhys.64.1045}%
  \BibitemOpen
  \bibfield  {author} {\bibinfo {author} {\bibfnamefont {M.~C.}\ \bibnamefont
  {Payne}}, \bibinfo {author} {\bibfnamefont {M.~P.}\ \bibnamefont {Teter}},
  \bibinfo {author} {\bibfnamefont {D.~C.}\ \bibnamefont {Allan}}, \bibinfo
  {author} {\bibfnamefont {T.~A.}\ \bibnamefont {Arias}}, \ and\ \bibinfo
  {author} {\bibfnamefont {J.~D.}\ \bibnamefont {Joannopoulos}},\ }\href
  {\doibase 10.1103/RevModPhys.64.1045} {\bibfield  {journal} {\bibinfo
  {journal} {Rev. Mod. Phys.}\ }\textbf {\bibinfo {volume} {64}},\ \bibinfo
  {pages} {1045} (\bibinfo {year} {1992})}\BibitemShut {NoStop}%
\bibitem [{\citenamefont {Vaugier}\ \emph {et~al.}(2012)\citenamefont
  {Vaugier}, \citenamefont {Jiang},\ and\ \citenamefont
  {Biermann}}]{PhysRevB.86.165105}%
  \BibitemOpen
  \bibfield  {author} {\bibinfo {author} {\bibfnamefont {L.}~\bibnamefont
  {Vaugier}}, \bibinfo {author} {\bibfnamefont {H.}~\bibnamefont {Jiang}}, \
  and\ \bibinfo {author} {\bibfnamefont {S.}~\bibnamefont {Biermann}},\ }\href
  {\doibase 10.1103/PhysRevB.86.165105} {\bibfield  {journal} {\bibinfo
  {journal} {Phys. Rev. B}\ }\textbf {\bibinfo {volume} {86}},\ \bibinfo
  {pages} {165105} (\bibinfo {year} {2012})}\BibitemShut {NoStop}%
\bibitem [{\citenamefont {Kaltak}(2015)}]{Merzuk2015}%
  \BibitemOpen
  \bibfield  {author} {\bibinfo {author} {\bibfnamefont {M.}~\bibnamefont
  {Kaltak}},\ }\emph {\bibinfo {title} {Merging GW with DMFT}},\ \href
  {http://othes.univie.ac.at/38099/} {Ph.D. thesis},\ \bibinfo  {school}
  {University of Vienna} (\bibinfo {year} {2015})\BibitemShut {NoStop}%
\bibitem [{\citenamefont {Liu}\ \emph {et~al.}(2018)\citenamefont {Liu},
  \citenamefont {Kim}, \citenamefont {Chen}, \citenamefont {Sarma},
  \citenamefont {Kresse},\ and\ \citenamefont
  {Franchini}}]{PhysRevMaterials.2.075003}%
  \BibitemOpen
  \bibfield  {author} {\bibinfo {author} {\bibfnamefont {P.}~\bibnamefont
  {Liu}}, \bibinfo {author} {\bibfnamefont {B.}~\bibnamefont {Kim}}, \bibinfo
  {author} {\bibfnamefont {X.-Q.}\ \bibnamefont {Chen}}, \bibinfo {author}
  {\bibfnamefont {D.~D.}\ \bibnamefont {Sarma}}, \bibinfo {author}
  {\bibfnamefont {G.}~\bibnamefont {Kresse}}, \ and\ \bibinfo {author}
  {\bibfnamefont {C.}~\bibnamefont {Franchini}},\ }\href {\doibase
  10.1103/PhysRevMaterials.2.075003} {\bibfield  {journal} {\bibinfo  {journal}
  {Phys. Rev. Materials}\ }\textbf {\bibinfo {volume} {2}},\ \bibinfo {pages}
  {075003} (\bibinfo {year} {2018})}\BibitemShut {NoStop}%
\bibitem [{\citenamefont {Zunger}\ \emph {et~al.}(1990)\citenamefont {Zunger},
  \citenamefont {Wei}, \citenamefont {Ferreira},\ and\ \citenamefont
  {Bernard}}]{PhysRevLett.65.353}%
  \BibitemOpen
  \bibfield  {author} {\bibinfo {author} {\bibfnamefont {A.}~\bibnamefont
  {Zunger}}, \bibinfo {author} {\bibfnamefont {S.-H.}\ \bibnamefont {Wei}},
  \bibinfo {author} {\bibfnamefont {L.~G.}\ \bibnamefont {Ferreira}}, \ and\
  \bibinfo {author} {\bibfnamefont {J.~E.}\ \bibnamefont {Bernard}},\ }\href
  {\doibase 10.1103/PhysRevLett.65.353} {\bibfield  {journal} {\bibinfo
  {journal} {Phys. Rev. Lett.}\ }\textbf {\bibinfo {volume} {65}},\ \bibinfo
  {pages} {353} (\bibinfo {year} {1990})}\BibitemShut {NoStop}%
\bibitem [{\citenamefont {van~de Walle}\ \emph {et~al.}(2002)\citenamefont
  {van~de Walle}, \citenamefont {Asta},\ and\ \citenamefont
  {Ceder}}]{VANDEWALLE2002539}%
  \BibitemOpen
  \bibfield  {author} {\bibinfo {author} {\bibfnamefont {A.}~\bibnamefont
  {van~de Walle}}, \bibinfo {author} {\bibfnamefont {M.}~\bibnamefont {Asta}},
  \ and\ \bibinfo {author} {\bibfnamefont {G.}~\bibnamefont {Ceder}},\ }\href
  {\doibase https://doi.org/10.1016/S0364-5916(02)80006-2} {\bibfield
  {journal} {\bibinfo  {journal} {Calphad}\ }\textbf {\bibinfo {volume} {26}},\
  \bibinfo {pages} {539 } (\bibinfo {year} {2002})}\BibitemShut {NoStop}%
\bibitem [{\citenamefont {van~de Walle}\ \emph {et~al.}(2013)\citenamefont
  {van~de Walle}, \citenamefont {Tiwary}, \citenamefont {de~Jong},
  \citenamefont {Olmsted}, \citenamefont {Asta}, \citenamefont {Dick},
  \citenamefont {Shin}, \citenamefont {Wang}, \citenamefont {Chen},\ and\
  \citenamefont {Liu}}]{VANDEWALLE201313}%
  \BibitemOpen
  \bibfield  {author} {\bibinfo {author} {\bibfnamefont {A.}~\bibnamefont
  {van~de Walle}}, \bibinfo {author} {\bibfnamefont {P.}~\bibnamefont
  {Tiwary}}, \bibinfo {author} {\bibfnamefont {M.}~\bibnamefont {de~Jong}},
  \bibinfo {author} {\bibfnamefont {D.}~\bibnamefont {Olmsted}}, \bibinfo
  {author} {\bibfnamefont {M.}~\bibnamefont {Asta}}, \bibinfo {author}
  {\bibfnamefont {A.}~\bibnamefont {Dick}}, \bibinfo {author} {\bibfnamefont
  {D.}~\bibnamefont {Shin}}, \bibinfo {author} {\bibfnamefont {Y.}~\bibnamefont
  {Wang}}, \bibinfo {author} {\bibfnamefont {L.-Q.}\ \bibnamefont {Chen}}, \
  and\ \bibinfo {author} {\bibfnamefont {Z.-K.}\ \bibnamefont {Liu}},\ }\href
  {\doibase https://doi.org/10.1016/j.calphad.2013.06.006} {\bibfield
  {journal} {\bibinfo  {journal} {Calphad}\ }\textbf {\bibinfo {volume} {42}},\
  \bibinfo {pages} {13 } (\bibinfo {year} {2013})}\BibitemShut {NoStop}%
\bibitem [{\citenamefont {Togo}\ \emph {et~al.}(2008)\citenamefont {Togo},
  \citenamefont {Oba},\ and\ \citenamefont {Tanaka}}]{PhysRevB.78.134106}%
  \BibitemOpen
  \bibfield  {author} {\bibinfo {author} {\bibfnamefont {A.}~\bibnamefont
  {Togo}}, \bibinfo {author} {\bibfnamefont {F.}~\bibnamefont {Oba}}, \ and\
  \bibinfo {author} {\bibfnamefont {I.}~\bibnamefont {Tanaka}},\ }\href
  {\doibase 10.1103/PhysRevB.78.134106} {\bibfield  {journal} {\bibinfo
  {journal} {Phys. Rev. B}\ }\textbf {\bibinfo {volume} {78}},\ \bibinfo
  {pages} {134106} (\bibinfo {year} {2008})}\BibitemShut {NoStop}%
\bibitem [{\citenamefont {Kronm\"{u}ller}\ and\ \citenamefont
  {Parkin}(2007)}]{Helmut2007}%
  \BibitemOpen
  \bibfield  {author} {\bibinfo {author} {\bibfnamefont {H.}~\bibnamefont
  {Kronm\"{u}ller}}\ and\ \bibinfo {author} {\bibfnamefont {S.}~\bibnamefont
  {Parkin}},\ }\href@noop {} {\emph {\bibinfo {title} {\textit{Handbook of
  magnetism and advanced magnetic materials}}}}\ (\bibinfo  {publisher}
  {Wiley},\ \bibinfo {year} {2007})\BibitemShut {NoStop}%
\bibitem [{\citenamefont {Benedek}\ and\ \citenamefont
  {Fennie}(2013)}]{Benedek2013}%
  \BibitemOpen
  \bibfield  {author} {\bibinfo {author} {\bibfnamefont {N.~A.}\ \bibnamefont
  {Benedek}}\ and\ \bibinfo {author} {\bibfnamefont {C.~J.}\ \bibnamefont
  {Fennie}},\ }\href
  {https://pubs.acs.org/action/downloadCitation?doi=10.1021/jp402046t&include=cit&format=ris&direct=true&downloadFileName=jp402046t}
  {\bibfield  {journal} {\bibinfo  {journal} {The Journal of Physical Chemistry
  C}\ }\textbf {\bibinfo {volume} {117}},\ \bibinfo {pages} {13339} (\bibinfo
  {year} {2013})}\BibitemShut {NoStop}%
\bibitem [{\citenamefont {Carsten}(2012)}]{Schinzer2012}%
  \BibitemOpen
  \bibfield  {author} {\bibinfo {author} {\bibfnamefont {S.}~\bibnamefont
  {Carsten}},\ }\href
  {http://www.ccp14.ac.uk/ccp/web-mirrors/pki/uni/pki/members/schinzer/stru_chem/perov/di_gold.html}
  {\emph {\bibinfo {title} {\textit{Distortion of Perovskites}}}}\ (\bibinfo
  {year} {2012})\BibitemShut {NoStop}%
\bibitem [{\citenamefont {Miller}\ and\ \citenamefont
  {Love}(1967)}]{Miller1967}%
  \BibitemOpen
  \bibfield  {author} {\bibinfo {author} {\bibfnamefont {S.~C.}\ \bibnamefont
  {Miller}}\ and\ \bibinfo {author} {\bibfnamefont {W.~F.}\ \bibnamefont
  {Love}},\ }\href@noop {} {\emph {\bibinfo {title} {\textit{Tables of
  Irreducible Representations of Space Groups and Co-representations of
  Magnetic Space Groups}}}}\ (\bibinfo  {publisher} {Pruett: Boulder},\
  \bibinfo {year} {1967})\BibitemShut {NoStop}%
\bibitem [{\citenamefont {Glazer}(1972)}]{Glazer1972}%
  \BibitemOpen
  \bibfield  {author} {\bibinfo {author} {\bibfnamefont {A.~M.}\ \bibnamefont
  {Glazer}},\ }\href {\doibase 10.1107/S0567740872007976} {\bibfield  {journal}
  {\bibinfo  {journal} {Acta Crystallographica Section B}\ }\textbf {\bibinfo
  {volume} {28}},\ \bibinfo {pages} {3384} (\bibinfo {year}
  {1972})}\BibitemShut {NoStop}%
\bibitem [{\citenamefont {Momma}\ and\ \citenamefont
  {Izumi}(2011)}]{Momma:db5098}%
  \BibitemOpen
  \bibfield  {author} {\bibinfo {author} {\bibfnamefont {K.}~\bibnamefont
  {Momma}}\ and\ \bibinfo {author} {\bibfnamefont {F.}~\bibnamefont {Izumi}},\
  }\href {\doibase 10.1107/S0021889811038970} {\bibfield  {journal} {\bibinfo
  {journal} {Journal of Applied Crystallography}\ }\textbf {\bibinfo {volume}
  {44}},\ \bibinfo {pages} {1272} (\bibinfo {year} {2011})}\BibitemShut
  {NoStop}%
\bibitem [{\citenamefont {Zhang}\ \emph {et~al.}(2018)\citenamefont {Zhang},
  \citenamefont {Gong}, \citenamefont {Li}, \citenamefont {Lin}, \citenamefont
  {Yan}, \citenamefont {Dong},\ and\ \citenamefont
  {Liu}}]{doi:10.1002/pssr.201800396}%
  \BibitemOpen
  \bibfield  {author} {\bibinfo {author} {\bibfnamefont {Y.}~\bibnamefont
  {Zhang}}, \bibinfo {author} {\bibfnamefont {J.}~\bibnamefont {Gong}},
  \bibinfo {author} {\bibfnamefont {C.}~\bibnamefont {Li}}, \bibinfo {author}
  {\bibfnamefont {L.}~\bibnamefont {Lin}}, \bibinfo {author} {\bibfnamefont
  {Z.}~\bibnamefont {Yan}}, \bibinfo {author} {\bibfnamefont {S.}~\bibnamefont
  {Dong}}, \ and\ \bibinfo {author} {\bibfnamefont {J.-M.}\ \bibnamefont
  {Liu}},\ }\href {\doibase 10.1002/pssr.201800396} {\bibfield  {journal}
  {\bibinfo  {journal} {physica status solidi (RRL) Rapid Research Letters}\
  }\textbf {\bibinfo {volume} {12}},\ \bibinfo {pages} {1800396} (\bibinfo
  {year} {2018})}\BibitemShut {NoStop}%
\bibitem [{\citenamefont {Paredes~Aulestia}\ \emph {et~al.}(2018)\citenamefont
  {Paredes~Aulestia}, \citenamefont {Cheung}, \citenamefont {Fang},
  \citenamefont {He}, \citenamefont {Yamaura}, \citenamefont {Lai},
  \citenamefont {Goh},\ and\ \citenamefont {Chen}}]{Paredes2018}%
  \BibitemOpen
  \bibfield  {author} {\bibinfo {author} {\bibfnamefont {E.~I.}\ \bibnamefont
  {Paredes~Aulestia}}, \bibinfo {author} {\bibfnamefont {Y.~W.}\ \bibnamefont
  {Cheung}}, \bibinfo {author} {\bibfnamefont {Y.-W.}\ \bibnamefont {Fang}},
  \bibinfo {author} {\bibfnamefont {J.}~\bibnamefont {He}}, \bibinfo {author}
  {\bibfnamefont {K.}~\bibnamefont {Yamaura}}, \bibinfo {author} {\bibfnamefont
  {K.~T.}\ \bibnamefont {Lai}}, \bibinfo {author} {\bibfnamefont {S.~K.}\
  \bibnamefont {Goh}}, \ and\ \bibinfo {author} {\bibfnamefont
  {H.}~\bibnamefont {Chen}},\ }\href {\doibase 10.1063/1.5035133} {\bibfield
  {journal} {\bibinfo  {journal} {Applied Physics Letters}\ }\textbf {\bibinfo
  {volume} {113}},\ \bibinfo {pages} {012902} (\bibinfo {year}
  {2018})}\BibitemShut {NoStop}%
\bibitem [{\citenamefont {Ekholm}\ \emph {et~al.}(2018)\citenamefont {Ekholm},
  \citenamefont {Gambino}, \citenamefont {J\"onsson}, \citenamefont
  {Tasn\'adi}, \citenamefont {Alling},\ and\ \citenamefont
  {Abrikosov}}]{PhysRevB.98.094413}%
  \BibitemOpen
  \bibfield  {author} {\bibinfo {author} {\bibfnamefont {M.}~\bibnamefont
  {Ekholm}}, \bibinfo {author} {\bibfnamefont {D.}~\bibnamefont {Gambino}},
  \bibinfo {author} {\bibfnamefont {H.~J.~M.}\ \bibnamefont {J\"onsson}},
  \bibinfo {author} {\bibfnamefont {F.}~\bibnamefont {Tasn\'adi}}, \bibinfo
  {author} {\bibfnamefont {B.}~\bibnamefont {Alling}}, \ and\ \bibinfo {author}
  {\bibfnamefont {I.~A.}\ \bibnamefont {Abrikosov}},\ }\href {\doibase
  10.1103/PhysRevB.98.094413} {\bibfield  {journal} {\bibinfo  {journal} {Phys.
  Rev. B}\ }\textbf {\bibinfo {volume} {98}},\ \bibinfo {pages} {094413}
  (\bibinfo {year} {2018})}\BibitemShut {NoStop}%
\bibitem [{\citenamefont {Fu}\ and\ \citenamefont
  {Singh}(2018)}]{PhysRevLett.121.207201}%
  \BibitemOpen
  \bibfield  {author} {\bibinfo {author} {\bibfnamefont {Y.}~\bibnamefont
  {Fu}}\ and\ \bibinfo {author} {\bibfnamefont {D.~J.}\ \bibnamefont {Singh}},\
  }\href {\doibase 10.1103/PhysRevLett.121.207201} {\bibfield  {journal}
  {\bibinfo  {journal} {Phys. Rev. Lett.}\ }\textbf {\bibinfo {volume} {121}},\
  \bibinfo {pages} {207201} (\bibinfo {year} {2018})}\BibitemShut {NoStop}%
\bibitem [{\citenamefont {He}\ and\ \citenamefont
  {Franchini}(2012)}]{PhysRevB.86.235117}%
  \BibitemOpen
  \bibfield  {author} {\bibinfo {author} {\bibfnamefont {J.}~\bibnamefont
  {He}}\ and\ \bibinfo {author} {\bibfnamefont {C.}~\bibnamefont {Franchini}},\
  }\href {\doibase 10.1103/PhysRevB.86.235117} {\bibfield  {journal} {\bibinfo
  {journal} {Phys. Rev. B}\ }\textbf {\bibinfo {volume} {86}},\ \bibinfo
  {pages} {235117} (\bibinfo {year} {2012})}\BibitemShut {NoStop}%
\bibitem [{\citenamefont {Liu}\ \emph {et~al.}(2019)\citenamefont {Liu},
  \citenamefont {Franchini}, \citenamefont {Marsman},\ and\ \citenamefont
  {Kresse}}]{Liu_2019}%
  \BibitemOpen
  \bibfield  {author} {\bibinfo {author} {\bibfnamefont {P.}~\bibnamefont
  {Liu}}, \bibinfo {author} {\bibfnamefont {C.}~\bibnamefont {Franchini}},
  \bibinfo {author} {\bibfnamefont {M.}~\bibnamefont {Marsman}}, \ and\
  \bibinfo {author} {\bibfnamefont {G.}~\bibnamefont {Kresse}},\ }\href
  {\doibase 10.1088/1361-648x/ab4150} {\bibfield  {journal} {\bibinfo
  {journal} {Journal of Physics: Condensed Matter}\ }\textbf {\bibinfo {volume}
  {32}},\ \bibinfo {pages} {015502} (\bibinfo {year} {2019})}\BibitemShut
  {NoStop}%
\bibitem [{\citenamefont {Ortenzi}\ \emph {et~al.}(2012)\citenamefont
  {Ortenzi}, \citenamefont {Mazin}, \citenamefont {Blaha},\ and\ \citenamefont
  {Boeri}}]{PhysRevB.86.064437}%
  \BibitemOpen
  \bibfield  {author} {\bibinfo {author} {\bibfnamefont {L.}~\bibnamefont
  {Ortenzi}}, \bibinfo {author} {\bibfnamefont {I.~I.}\ \bibnamefont {Mazin}},
  \bibinfo {author} {\bibfnamefont {P.}~\bibnamefont {Blaha}}, \ and\ \bibinfo
  {author} {\bibfnamefont {L.}~\bibnamefont {Boeri}},\ }\href {\doibase
  10.1103/PhysRevB.86.064437} {\bibfield  {journal} {\bibinfo  {journal} {Phys.
  Rev. B}\ }\textbf {\bibinfo {volume} {86}},\ \bibinfo {pages} {064437}
  (\bibinfo {year} {2012})}\BibitemShut {NoStop}%
\bibitem [{\citenamefont {Sangiovanni}\ \emph {et~al.}(2006)\citenamefont
  {Sangiovanni}, \citenamefont {Toschi}, \citenamefont {Koch}, \citenamefont
  {Held}, \citenamefont {Capone}, \citenamefont {Castellani}, \citenamefont
  {Gunnarsson}, \citenamefont {Mo}, \citenamefont {Allen}, \citenamefont {Kim},
  \citenamefont {Sekiyama}, \citenamefont {Yamasaki}, \citenamefont {Suga},\
  and\ \citenamefont {Metcalf}}]{PhysRevB.73.205121}%
  \BibitemOpen
  \bibfield  {author} {\bibinfo {author} {\bibfnamefont {G.}~\bibnamefont
  {Sangiovanni}}, \bibinfo {author} {\bibfnamefont {A.}~\bibnamefont {Toschi}},
  \bibinfo {author} {\bibfnamefont {E.}~\bibnamefont {Koch}}, \bibinfo {author}
  {\bibfnamefont {K.}~\bibnamefont {Held}}, \bibinfo {author} {\bibfnamefont
  {M.}~\bibnamefont {Capone}}, \bibinfo {author} {\bibfnamefont
  {C.}~\bibnamefont {Castellani}}, \bibinfo {author} {\bibfnamefont
  {O.}~\bibnamefont {Gunnarsson}}, \bibinfo {author} {\bibfnamefont {S.-K.}\
  \bibnamefont {Mo}}, \bibinfo {author} {\bibfnamefont {J.~W.}\ \bibnamefont
  {Allen}}, \bibinfo {author} {\bibfnamefont {H.-D.}\ \bibnamefont {Kim}},
  \bibinfo {author} {\bibfnamefont {A.}~\bibnamefont {Sekiyama}}, \bibinfo
  {author} {\bibfnamefont {A.}~\bibnamefont {Yamasaki}}, \bibinfo {author}
  {\bibfnamefont {S.}~\bibnamefont {Suga}}, \ and\ \bibinfo {author}
  {\bibfnamefont {P.}~\bibnamefont {Metcalf}},\ }\href {\doibase
  10.1103/PhysRevB.73.205121} {\bibfield  {journal} {\bibinfo  {journal} {Phys.
  Rev. B}\ }\textbf {\bibinfo {volume} {73}},\ \bibinfo {pages} {205121}
  (\bibinfo {year} {2006})}\BibitemShut {NoStop}%
\bibitem [{\citenamefont {Varignon}\ \emph {et~al.}(2019)\citenamefont
  {Varignon}, \citenamefont {Bibes},\ and\ \citenamefont
  {Zunger}}]{Varignon2019}%
  \BibitemOpen
  \bibfield  {author} {\bibinfo {author} {\bibfnamefont {J.}~\bibnamefont
  {Varignon}}, \bibinfo {author} {\bibfnamefont {M.}~\bibnamefont {Bibes}}, \
  and\ \bibinfo {author} {\bibfnamefont {A.}~\bibnamefont {Zunger}},\ }\href
  {http://citation-needed.springer.com/v2/references/10.1038/s41467-019-09698-6?format=refman&flavour=citation}
  {\bibfield  {journal} {\bibinfo  {journal} {Nat Commun}\ }\textbf {\bibinfo
  {volume} {10}},\ \bibinfo {pages} {1658} (\bibinfo {year}
  {2019})}\BibitemShut {NoStop}%
\bibitem [{\citenamefont {Pittalis}\ \emph {et~al.}(2017)\citenamefont
  {Pittalis}, \citenamefont {Vignale},\ and\ \citenamefont
  {Eich}}]{PhysRevB.96.035141}%
  \BibitemOpen
  \bibfield  {author} {\bibinfo {author} {\bibfnamefont {S.}~\bibnamefont
  {Pittalis}}, \bibinfo {author} {\bibfnamefont {G.}~\bibnamefont {Vignale}}, \
  and\ \bibinfo {author} {\bibfnamefont {F.~G.}\ \bibnamefont {Eich}},\ }\href
  {\doibase 10.1103/PhysRevB.96.035141} {\bibfield  {journal} {\bibinfo
  {journal} {Phys. Rev. B}\ }\textbf {\bibinfo {volume} {96}},\ \bibinfo
  {pages} {035141} (\bibinfo {year} {2017})}\BibitemShut {NoStop}%
\bibitem [{\citenamefont {Sharma}\ \emph {et~al.}(2018)\citenamefont {Sharma},
  \citenamefont {Gross}, \citenamefont {Sanna},\ and\ \citenamefont
  {Dewhurst}}]{Sharma2018}%
  \BibitemOpen
  \bibfield  {author} {\bibinfo {author} {\bibfnamefont {S.}~\bibnamefont
  {Sharma}}, \bibinfo {author} {\bibfnamefont {E.~K.~U.}\ \bibnamefont
  {Gross}}, \bibinfo {author} {\bibfnamefont {A.}~\bibnamefont {Sanna}}, \ and\
  \bibinfo {author} {\bibfnamefont {J.~K.}\ \bibnamefont {Dewhurst}},\ }\href
  {\doibase 10.1021/acs.jctc.7b01049} {\bibfield  {journal} {\bibinfo
  {journal} {Journal of Chemical Theory and Computation}\ }\textbf {\bibinfo
  {volume} {14}},\ \bibinfo {pages} {1247} (\bibinfo {year}
  {2018})}\BibitemShut {NoStop}%
\bibitem [{\citenamefont {Triebl}\ \emph {et~al.}(2018)\citenamefont {Triebl},
  \citenamefont {Kraberger}, \citenamefont {Mravlje},\ and\ \citenamefont
  {Aichhorn}}]{PhysRevB.98.205128}%
  \BibitemOpen
  \bibfield  {author} {\bibinfo {author} {\bibfnamefont {R.}~\bibnamefont
  {Triebl}}, \bibinfo {author} {\bibfnamefont {G.~J.}\ \bibnamefont
  {Kraberger}}, \bibinfo {author} {\bibfnamefont {J.}~\bibnamefont {Mravlje}},
  \ and\ \bibinfo {author} {\bibfnamefont {M.}~\bibnamefont {Aichhorn}},\
  }\href {\doibase 10.1103/PhysRevB.98.205128} {\bibfield  {journal} {\bibinfo
  {journal} {Phys. Rev. B}\ }\textbf {\bibinfo {volume} {98}},\ \bibinfo
  {pages} {205128} (\bibinfo {year} {2018})}\BibitemShut {NoStop}%
\bibitem [{\citenamefont {Zhang}\ \emph {et~al.}(2017)\citenamefont {Zhang},
  \citenamefont {Sun}, \citenamefont {Perdew},\ and\ \citenamefont
  {Wu}}]{PhysRevB.96.035143}%
  \BibitemOpen
  \bibfield  {author} {\bibinfo {author} {\bibfnamefont {Y.}~\bibnamefont
  {Zhang}}, \bibinfo {author} {\bibfnamefont {J.}~\bibnamefont {Sun}}, \bibinfo
  {author} {\bibfnamefont {J.~P.}\ \bibnamefont {Perdew}}, \ and\ \bibinfo
  {author} {\bibfnamefont {X.}~\bibnamefont {Wu}},\ }\href {\doibase
  10.1103/PhysRevB.96.035143} {\bibfield  {journal} {\bibinfo  {journal} {Phys.
  Rev. B}\ }\textbf {\bibinfo {volume} {96}},\ \bibinfo {pages} {035143}
  (\bibinfo {year} {2017})}\BibitemShut {NoStop}%
\bibitem [{\citenamefont {Popescu}\ and\ \citenamefont
  {Zunger}(2010)}]{PhysRevLett.104.236403}%
  \BibitemOpen
  \bibfield  {author} {\bibinfo {author} {\bibfnamefont {V.}~\bibnamefont
  {Popescu}}\ and\ \bibinfo {author} {\bibfnamefont {A.}~\bibnamefont
  {Zunger}},\ }\href {\doibase 10.1103/PhysRevLett.104.236403} {\bibfield
  {journal} {\bibinfo  {journal} {Phys. Rev. Lett.}\ }\textbf {\bibinfo
  {volume} {104}},\ \bibinfo {pages} {236403} (\bibinfo {year}
  {2010})}\BibitemShut {NoStop}%
\bibitem [{\citenamefont {Liu}\ \emph {et~al.}(2016)\citenamefont {Liu},
  \citenamefont {Reticcioli}, \citenamefont {Kim}, \citenamefont {Continenza},
  \citenamefont {Kresse}, \citenamefont {Sarma}, \citenamefont {Chen},\ and\
  \citenamefont {Franchini}}]{PhysRevB.94.195145}%
  \BibitemOpen
  \bibfield  {author} {\bibinfo {author} {\bibfnamefont {P.}~\bibnamefont
  {Liu}}, \bibinfo {author} {\bibfnamefont {M.}~\bibnamefont {Reticcioli}},
  \bibinfo {author} {\bibfnamefont {B.}~\bibnamefont {Kim}}, \bibinfo {author}
  {\bibfnamefont {A.}~\bibnamefont {Continenza}}, \bibinfo {author}
  {\bibfnamefont {G.}~\bibnamefont {Kresse}}, \bibinfo {author} {\bibfnamefont
  {D.~D.}\ \bibnamefont {Sarma}}, \bibinfo {author} {\bibfnamefont {X.-Q.}\
  \bibnamefont {Chen}}, \ and\ \bibinfo {author} {\bibfnamefont
  {C.}~\bibnamefont {Franchini}},\ }\href {\doibase 10.1103/PhysRevB.94.195145}
  {\bibfield  {journal} {\bibinfo  {journal} {Phys. Rev. B}\ }\textbf {\bibinfo
  {volume} {94}},\ \bibinfo {pages} {195145} (\bibinfo {year}
  {2016})}\BibitemShut {NoStop}%
\bibitem [{\citenamefont {Birch}(1947)}]{PhysRev.71.809}%
  \BibitemOpen
  \bibfield  {author} {\bibinfo {author} {\bibfnamefont {F.}~\bibnamefont
  {Birch}},\ }\href {\doibase 10.1103/PhysRev.71.809} {\bibfield  {journal}
  {\bibinfo  {journal} {Phys. Rev.}\ }\textbf {\bibinfo {volume} {71}},\
  \bibinfo {pages} {809} (\bibinfo {year} {1947})}\BibitemShut {NoStop}%
\end{thebibliography}%

\end{document}